\documentclass{osa-article}

\usepackage{footnote}
\usepackage{amsmath}
\usepackage{braket}
\usepackage{graphicx}
\usepackage{epstopdf}
\usepackage{upgreek}
\usepackage{bm}
\usepackage{import}
\usepackage{endnotes}
\usepackage{hyperref}
\usepackage{nicefrac}

\begin{document}

\title{Ultrafast quantum dynamics driven by the strong space charge field of a relativistic electron beam}

\author{D.~Cesar\authormark{1,*}, 
A.~Acharya\authormark{1}, 
J.~P.~Cryan\authormark{1,2}, 
A.Kartsev\authormark{1,3}, 
M.F.~Kling\authormark{1,2,4}, 
A.M.~Lindenberg\authormark{1,2,5,8}, 
C.D.~Pemmaraju\authormark{8,9}, 
A.D.~Poletayev\authormark{1,5,8,10},
V.S.~Yakovlev\authormark{6,7}, 
A.~Marinelli\authormark{1}}

\authormark{1}SLAC National Accelerator Laboratory, Menlo Park, CA 94025, USA\\
\authormark{2}Stanford PULSE Institute, SLAC National Accelerator Laboratory, Menlo Park, CA, USA.\\
\authormark{3}Bauman Moscow State Technical University, Moscow, 105005, Russia\\
\authormark{4}Department of Applied Physics, Stanford University, Stanford, California 94305, USA\\
\authormark{5}Department of Materials Science and Engineering, Stanford University, Stanford, California 94305, USA\\
\authormark{6}Max Planck Institute of Quantum Optics, 85748 Garching, Germany\\
\authormark{7}Department of Physics, Ludwig-Maximilians-Universit\"{a}t Munich, 85748 Garching, Germany\\
\authormark{8}Stanford Institute for Materials and Energy Sciences, SLAC National Laboratory, Menlo Park, CA, USA\\
\authormark{9}Theory Institute for Materials and Energy Spectroscopies, SLAC National Accelerator Laboratory, Menlo Park, California 94025, USA\\
\authormark{10}Present Address: Department of Materials, University of Oxford, Oxford OX13RQ, UK\\
\email{\authormark{*}dcesar@slac.stanford.edu}

\begin{abstract}
In this article, we illustrate how the Coulomb field of a highly relativistic electron beam can be shaped into a broadband pulse suitable for driving ultrafast and strong-field physics. In contrast to a solid-state laser, the Coulomb field creates a pulse which can be intrinsically synchronized with an x-ray free electron laser (XFEL), can have a cutoff frequency which is broadly tunable from THz to EUV, and which acts on target systems as a ``half-cycle'' impulse. Explicit examples are presented to emphasize how the unique features of this excitation can be a tool for novel science at XFEL facilities like the LCLS.
\end{abstract}


\section*{Introduction} 


Over the last decade, x-ray free electron lasers (XFELs) have been a transformative tool for the physical sciences by delivering intense bursts of x-rays which can probe electronic structure with site-specificity and ultrafast temporal resolution\,\cite{bostedt_linac_2016}. The high intensity of the XFEL allows the pump-probe framework to be applied to traditional x-ray imaging and spectroscopy, where the pump is most often either a secondary x-ray pulse\,\cite{lutman_experimental_2013, marinelli_high-intensity_2015, lutman_fresh-slice_2016} or a solid-state laser that has been synchronized to the x-ray probe \cite{petrovic_transient_2012,cryan_auger_2010,glownia_time-resolved_2010}. More recently, novel electron-beam shaping techniques have extended XFEL temporal resolution to the attosecond regime and made it possible to chart coherent charge dynamics in molecular systems\,\cite{duris_tunable_2020, li_attosecond_2022}. 

\begin{figure}
    \centering
    \includegraphics[scale=0.8]{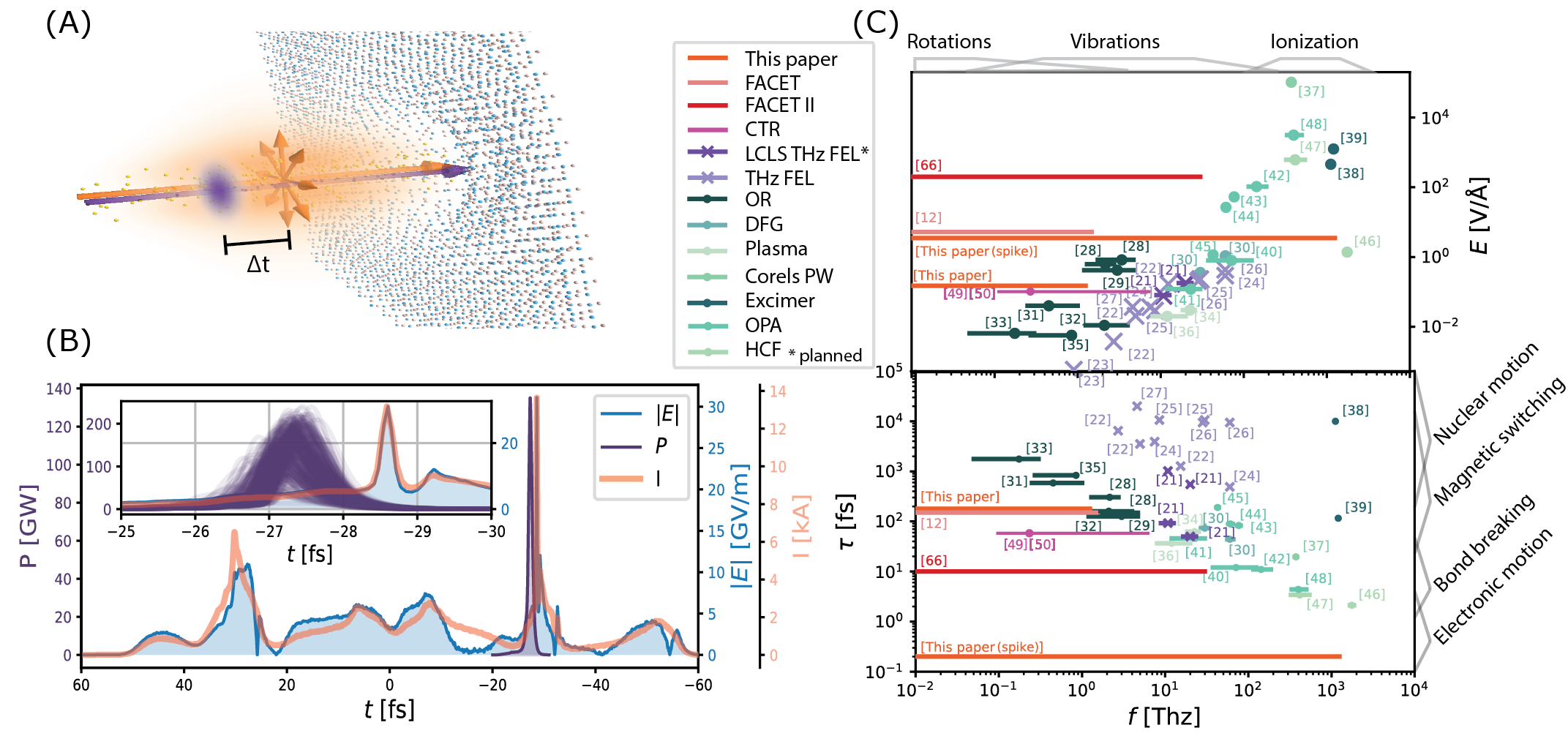}
    \caption{Parameter space for Coulomb driven quantum dynamics. (A) Cartoon of a space-charge pump, x-ray probe experiment. (B) The space-charge field (blue), electron beam current (red) and the power of sample XFEL pulses (purple) from start-to-end simulations of the LCLS-II cuS beamline. (C) Parameter space plots comparing conventional FEL\,\cite{zhang_high-power_2020,koevener_thz_2016,ozerov_thz_2017,bakker_intensity-resolved_2010,li_heting__2021,ortega_extension_2006,kawase_extremely_2020} and laser sources (OR: optical rectification, DFG: difference frequency generation, PW: petawatt, OPA: optical parametric amplification, HCF: hollow core fiber) \,\cite{shalaby_demonstration_2015,vicario_generation_2014,sell_phase-locked_2008,wu_highly_2018,hauri_strong-field_2011,fulop_efficient_2014,oh_intense_2013,fulop_highly_2016,dey_highly_2017,yoon_realization_2021,shaw_ultrahigh-brightness_1993,mizoguchi_100-fs_1992,liang_high-energy_2017,junginger_single-cycle_2010,deng_carrier-envelope-phase-stable_2012,andriukaitis_90_2011,grafenstein_multi-millijoule_2020,elu_table-top_2019,travers_high-energy_2019,ouille_relativistic-intensity_2020,rivas_next_2017} to beam-driven fields (including CTR: coherent transition radiation\,\cite{daranciang_single-cycle_2011,wu_intense_2013}). The references follow those of topical reviews\,\cite{fulop_laser-driven_2020,hzdr_thz_2022,herziger_33_2007}. Notice that the conventional sources appear to lie along lines (in the log-log plot) because increasing frequency implies a smaller diffraction-limited spot size\,\cite{focusing_footnote}. The space-charge fields violate this trend because an electron beam can be focused to sub-wavelength spot-sizes (here we chose 20\,$\upmu$m rms as a compromise between pumped-volume and field strength). The resulting beam driven fields have a wider bandwidth (FWHM in intensity), shorter duration (FWHM in intensity), and larger field strength (peak) than their conventional counterparts. }
    \label{fig:cartoon} 
\end{figure}

In this article, we propose to extend the reach of ultrafast science at XFELs by directly pumping quantum systems with the same relativistic electron beam used to generate x-rays (Fig.\,\ref{fig:cartoon}a). The beam interacts with valence electrons primarily through the optical cross-section (rather than the comparatively small collisional cross-section\,\cite{seltzer_cross_1988}), such that we can think of its Coulomb field as a strong, ``half-cycle'', radially-polarized laser. Pioneering experiments at the Sub-Picosecond Pulse Source~(SPPS)\,\cite{krejcik_commissioning_2003} and FACET\,\cite{clarke_facet_nodate} facilities demonstrate that this beam field can be used to drive magnetic switching\,\cite{tudosa_ultimate_2004, lu_electronic_2019} and ultrafast changes in conductivity\,\cite{oshea_conductivity_2019}. 

More generally, the field of the beam is an ultra-broadband impulse which drives processes ranging from ``DC''-like Stark shifts up to ultraviolet ionization of valence electrons. The resulting dynamics are a complex superposition of excited states most similar to those created by single-cycle THz sources (Fig.\,\ref{fig:cartoon}b); however, we identify four main distinctions: Firstly, the pulse length can be much shorter than that of conventional lasers; indeed, by using techniques developed to produce attosecond x-rays\,\cite{zhang_double_2019,macarthur_phase-stable_2019}, we can create half-cycle fields with intensity profiles as short as 250\,as, corresponding to 12\,eV bandwidth; Secondly, the field is truly a ``unipolar'' impulse, such that the direct momentum transfer $A=-\int E dt$ is nonzero. Thirdly, we can create strong fields up to tens of V/\AA\, by focusing the electron beam to sub-wavelength spot sizes. And, lastly, the field is intrinsically synchronized to the x-ray pulses generated by the bunch.  

The importance of the last point should not be underestimated: state-of-the-art timing jitter between an optical laser and an XFEL pulse is between 20 and 100\,fs \cite{kang_hard_2017}, such that time-stamping techniques must be used to re-sort data on a shot-by-shot basis\,\cite{hartmann_sub-femtosecond_2014}---a feat which will become increasingly difficult as the next generation of high-rate superconducting FELs come online. Intrinsic synchronization bypasses this issue and directly enables attosecond pump-probe experiments. 

In this paper, we will discuss how to generate, characterize, and ultimately use ultrafast space-charge and x-ray pulses for photon-electron pump probe experiments~(PEPPEx). Our discussion is grounded by start-to-end simulations at the LCLS-II CuS beamline\,\cite{woodley_lcls_nodate}, although the same concepts are adaptable to other FEL beamlines. We show how the electron beam's phase space can be manipulated with a laser heater to produce two spikes suitable for a pump-probe experiment. We calculate the x-ray and space-charge fields from this beam, and we show how  photo-electron streaking can be used to reconstruct those same fields. Then we discuss the interaction of the space-charge field with three distinct quantum systems.

\section{Pulse synthesis}
To prepare an ultrafast pump-probe experiment we first shape the electron beam to have two short current spikes: the later (`tail') spike is used to generate x-rays, while the earlier (`head') spike will be used for its Coulomb field. A small magnetic chicane can then be used to delay the electron beam relative to the x-rays and precisely set the overlap between the Coulomb field and the x-ray pulse. To optimize this arrangement, it is necessary to control the chirp of the two current spikes separately, since the Coulomb spike should be fully compressed while the lasing spike should be strongly chirped (the addition of an x-ray delay line would create additional flexibility, but we do not use one here).

\begin{figure}
    \centering
    \includegraphics[scale=0.75]{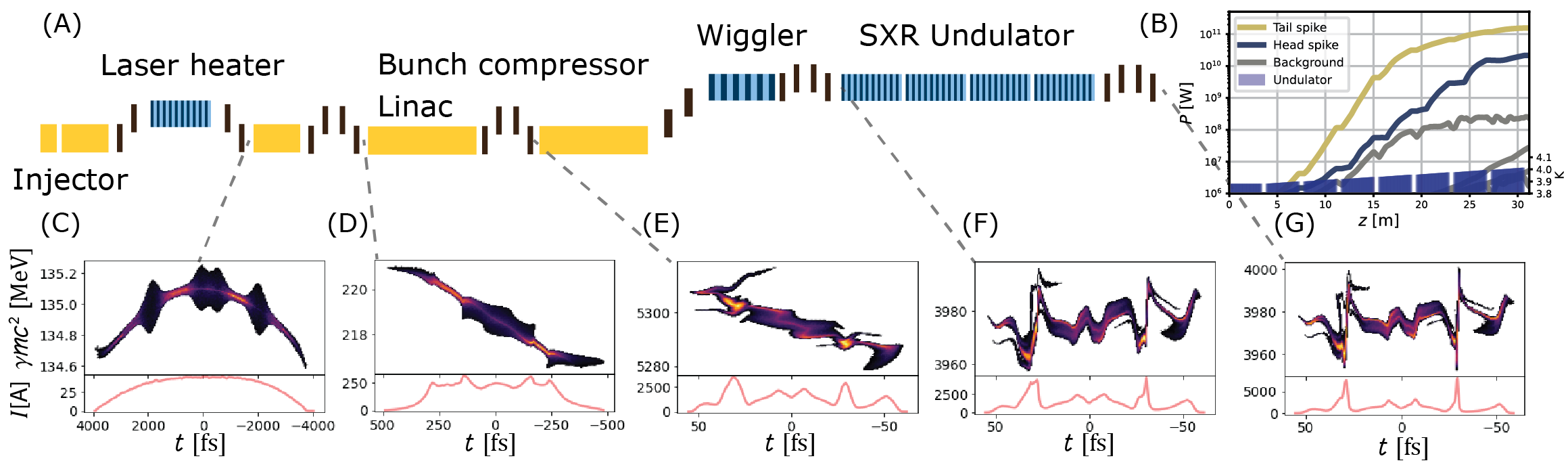}
    \caption{(A) Cartoon of the LCLS-II cuS beamline. (C-G) Snapshots of the electron beam longitudinal phase space. A coherent energy modulation is applied by a stack of 0.7\,ps long laser-heater pulses. This modulation evolves into two large spikes separated by approximately 50\,fs. The undulator (450\,eV) is chirp-taper matched to the tail (left) spike so that the (ensemble averaged) peak power (B) from the tail grows fastest and dominates the total x-ray radiation.}
    \label{fig:pulse_synthesis}
\end{figure}

One method to create such a bunch is based on laser heater shaping, as described in detail in ref\,\cite{cesar_electron_2021}. In short, the process plays out as illustrated in Fig.\,\ref{fig:pulse_synthesis}, which is based on start-to-end simulations of the LCLS-II CuS beamline\,\cite{woodley_lcls_nodate} (using IMPACT-T for the injector and Elegant for the transport from the injector to the undulators\,\cite{qiang_start--end_2017,wang_benchmark_2015,borland_elegant_2000}). First, a stack of Gaussian laser heater pulses creates time-dependent slice-energy spread. Then, after the first bunch compressor, the energy spread is converted into current modulations which seed the microbunching instability and are amplified during transport. Finally, anomalous dispersion in the magnetic dogleg compresses the space-charge induced chirp into two large current spikes. The compression ratio of the two spikes can be controlled separately by changing the delay between successive heater pulses (relative both to each other and to the beam center) in order to optimize the microbunching gain for each spike separately. 

This procedure leaves us with a strongly chirped tail spike which we use to generate the x-ray probe by matching the undulator taper to the beam chirp \,\cite{zhang_double_2019,duris_tunable_2020}. Genesis\,\cite{reiche_genesis_1999} simulations based on the LCLS-II soft x-ray line\,\cite{woodley_lcls_nodate} show that, with a matched taper, the tail spike can produce 450\,eV soft x-rays with an average power over 120\,GW, compared to $<10$\,GW from the head spike. The inset to Fig.\,\ref{fig:cartoon}(B) shows many individual simulations of the tail spike, each with a unique random seed leading to unique FEL dynamics. Most often the individual pulses have a 0.6\,fs full-width-at-half-maximum (FWHM) pulse length, but varying amounts of post-saturation slippage broaden the ensemble-averaged intensity to 0.8\,fs FWHM. This 0.2\,fs difference is expected to be the dominant component of the pump-probe jitter between the x-rays and the electron beam.

After lasing, a 40\,fs chicane delay overlaps the x-ray pulse with the current spike at the head of the bunch, as shown in Fig.\,\ref{fig:cartoon}(B). At the same time, this chicane fully compresses the head spike into a 13.5\,kA peak with a FWHM width of only 250\,as and an energy spread of 25\,MeV. Small changes in dispersion from the chicane can be used to control the pump-probe delay within an effective Rayleigh length of $\approx35$\,fs (outside of which the current spike begins to de-compress). Note also that the head spike looses a small, but variable, amount of energy while lasing, and this leads to negligible jitter (compared to rf induced jitter) in the compressed spike-width---the current profile in Fig.\,\ref{fig:cartoon}(B) shows a typical case.

Meanwhile, the electrons in our beam collectively form a strong electromagnetic impulse which closely approximates a ``half-cycle" laser pulse. For a transversely Gaussian charge density we can estimate the electric field as:
\begin{align}
\begin{split}
\label{eq:SC_round_gauss}
\rho_\perp &= \frac{I(t)}{2 \pi \sigma_r^2 } e^{-\frac{1}{2}\left(\frac{r}{\sigma_r}\right)^2} \\
\bm{E}(\bm{r},t) &= \hat{r} \frac{Z_0 I(t)}{2 \pi r}\left(1-e^{-\frac{1}{2}\left(\frac{r}{\sigma_r}\right)^2}\right)
\end{split}
\end{align}
with an associated azimuthal magnetic field given by $\bm{B}=(\bm{\beta}c\times\bm{E})$.  This model for the field produced by the beam is valid provided that: (a) we only consider frequencies commensurate with the bunch length $f	\lessapprox1/\sigma_t$, such that the field of the individual electrons add coherently and we can neglect stochastic effects\cite{robicheaux_coherent_2000,seltzer_cross_1988}; (b) we consider only cases in which the beam can be approximated by a long cylinder in its rest frame: $\sigma_r<\gamma c \sigma_t$; and finally (c) that the beam density is a 1D Gaussian cylinder. Of these conditions, only (c) is regularly violated in practice. In particular, while an individual slice of the electron beam may resemble a Gaussian cross section, the centroid is often a strong function of $t$ thanks to coherent synchrontron radiation (CSR) induced energy loss in the bunch compressor. Thus, to get an accurate space-charge field, we use the numerical Poisson solver from the tracking code GPT\,\cite{van_der_geer_general_nodate}. It is this field, evaluated a single point in space, that is plotted as the blue curve in Fig.\,\ref{fig:cartoon}(B).

\section {Pulse Characterization}
The ``gold standard" for the metrology of sub-femtosecond XUV and X-ray pulses is photoemission streaking with long wavelength fields.
In this technique, an ultrashort laser pulse is overlapped with a longer wavelength dressing laser field. 
The combined field is incident on a gas-sample and the momentum distribution of electrons ionized by the ultrashort field are displaced~(``streaked") by the long-wavelength dressing field ~\cite{orfanos_attosecond_2019,itatani_attosecond_2002,drescher_x-ray_2001}. 
This two-color ionization process encodes the information of the short-wavelength pulse into the measured photoelectron momentum distribution.
The pulse profiles can be retrieved from the resultant spectrogram~(recorded by scanning the relative delay between the x-ray pulse and IR field) via a number of proposed algorithms\,\cite{mairesse_frequency-resolved_2005,gagnon_accurate_2008,chini_characterizing_2010,keathley_volkov_2016}. 
To date, streaking at x-ray free electron laser~(XFEL) facilities relies on an external laser pulse which must be temporary overlapped with the XFEL pulse inside a gas phase sample.
The shot-to-shot jitter of the relative laser/x-ray arrival time precludes measurements which are not single-shot, and thus the dressing field has always been either long wavelength\,\cite{haynes_clocking_2021} or circularly polarized\,\cite{hartmann_attosecond_2018,heider_megahertz-compatible_2019, duris_tunable_2020}. 
But, by using a naturally synchronized, unipolar streaking field we have the possibility to average together many independent shots and thus greatly increase measurement sensitivity.

Streaked photo-electron spectra are conventionally calculated within the strong field approximation to the Schr\"{o}dinger equation for a  single active electron, which ignores the effect of the Coulomb potential on the emitted electron.
In this approximation, the probability for observing an electron with momentum $\bm{k}$ is given by:
\begin{align}
\label{eq:SFA}
W(\bm{k},\tau) &=\left|\int_{-\infty}^\infty{ dt\, e^{i(Q(t))}D(\bm{k+A(t)})\bm{E}_{x}(t-\tau)}\right|^2 \\
Q(t) &=\int_{-\infty}^{t}{dt'\left(\text{IP}+ \frac{1}{2}(\bm{k}+\bm{A}(t'))^2\right)}
\end{align}
where $D(\bm{k})$ is the photoionization dipole moment (here approximated with the value for hydrogen atoms $D(\bm{k})=\frac{\bm{k}}{(k^2+2\text{IP})^3}$), $\text{IP}$ is the ionization potential, $\bm{E}_{x}(t')$ is the electric field of the ionizing x-ray pulse, and $A(t)=-\int dt E(t)$ is vector potential of the streaking (space-charge) field with $A(\infty)=0$.
$Q(t)$ is the so-called Volkov phase. 
In anticipation of averaging over many shots, we will re-write this in terms of the first-order correlation function $\bm{\Gamma}(t_1,t_2)=\langle E(t_1)E^*(t_2)\rangle$ which allows us to account for the SASE jitter:
\begin{align}
\label{eq:SFA_average}
W(\bm{k},\tau) &=\int_{-\infty}^\infty{\int_{-\infty}^\infty{ dt_1dt_2\, e^{i(Q(t_1)-Q(t_2))}D(\bm{k+A(t_1)})D^*(\bm{k+A(t_2)})\bm{
\Gamma}(t_1-\tau,t_2-\tau)}} 
\end{align}

Within the quasi-classical model we imagine that an ionizing x-ray pulse creates photo-electrons at time $t_i$ with probability $I(t)$~(the intensity of the incident x-ray pulse) and energy $\nicefrac{k^2}{2}$ with probability $I(w)$ (possibly with some $t-\omega$ correlation due to chirp). If the streaking field is strong, then the Coulomb potential can be neglected and the now-free electrons are accelerated to a final momentum of $\bm{k}_f=\bm{k}_i-\bm{A}$.

\begin{figure}
    \centering
    \includegraphics[scale=0.73]{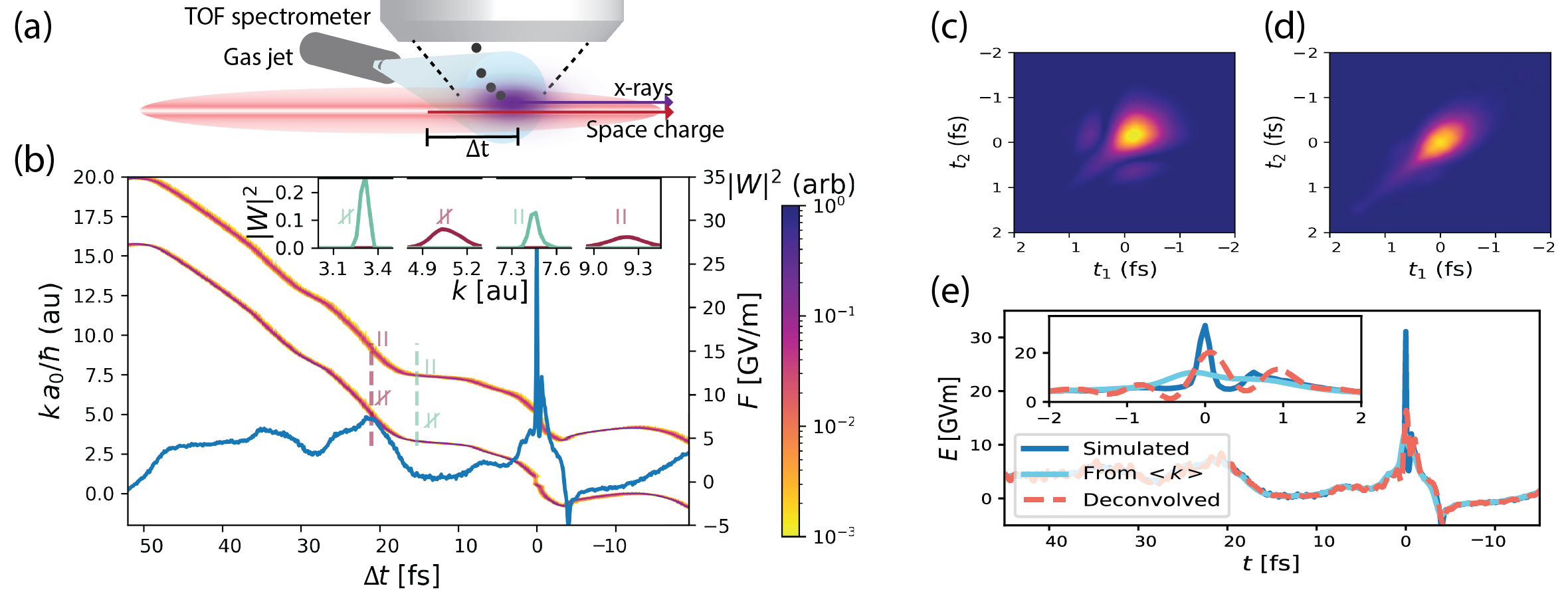}
    \caption{Ultrafast pulse metrology by photo-electron streaking. (a) Cartoon of the streaking measurement. (b) Spectrogram showing the photo-electron spectrum as a function of pump-probe delay. Also shown is the streaking field (blue). The inset shows the spectrum measured at two time delays as indicated by the dashed lines. (c) Simulated correlation function $\Gamma(t_1,t_2)$ (d) Reconstructed correlation function. (e) Simulated and reconstructed Coulomb fields}
    \label{fig:spectrogram}
\end{figure}

In Fig.\ref{fig:spectrogram}, we simulate a spectrogram $W(k,\tau)$ based on the fields from the previous section (for photo-electrons emitted from a single point in space and collected parallel to the x-ray polarization). For each time delay, we average the spectrum from equation \ref{eq:SFA} calculated using many independent random seeds for the SASE radiation (but neglecting any changes in $A$ due to machine jitter). The quasi-classical model allows a straightforward interpretation: we see two distinct populations separated by $2 k_i=2\sqrt{2(\hbar\omega-\text{IP})} $ for an ionization potential $\text{IP}$, corresponding to electrons emitted in directions parallel and anti-parallel to the streaking field. Each group of electrons then follows the vector $A(t)$ as the pump-probe delay is changed. Where the streaking field is large, the spectral width of the photo-electron peak is proportional to its pulse duration, and where the streaking field is weak it is proportional to the x-ray spectrum. In-between, as we can see in the inset to Fig.\,\ref{fig:spectrogram}, the width of the parallel and anti-parallel populations are different, due the the average chirp of the x-ray beam (originating from the chirp-taper FEL configuration \cite{duris_tunable_2020}).

After averaging over many shots to produce a spectrogram, we are no longer sensitive to the pulse length of individual shots. Instead, we measure the first order correlation function $\Gamma(t_1,t_2)$ , as indicated by Eq.\,\ref{eq:SFA_average}. An example $\Gamma(t_1,t_2)$ calculated directly from the FEL simulations is shown in Fig.\,\ref{fig:spectrogram}(b): the main diagonal ($t_1=t_2$) yields the average FEL power, while the off-diagonals ($t_1 \propto-t_2$) are related to the coherence length. Because the FEL is not a stationary process, the coherence length is not constant, but in fact increases towards the head (upper right) of the pulse, where the slippage has built up coherent power. 

In order to fully reconstruct $\Gamma(t_1,t_2)$ and the streaking field $A(t)$, we adopt an procedure similar to \cite{li_characterizing_2018} and \cite{keathley_volkov_2016}, in which we first estimate the streaking field based on the average spectrum $A(t)\approx \langle W(\tau) \rangle|_k$ and then iteratively solve the least squares problem for $\Gamma(t_1,t_2)$ (i.e. Eq.\,\ref{eq:SFA_average}). This avoids the approximations used in the common FROG-CRAB algorithms, which are not applicable to our case. To speed-up convergence, we represent the underlying $\Gamma$ in a compact basis, an extension of the Von Neumann basis used in\,\cite{li_characterizing_2018}. We can then refine our guess for $A(t)$ because we know that, within the quasi-classical approximation, the average spectrum is the convolution of the streaking field and the average x-ray power: $\langle W(\tau) \rangle|_k=A(\tau) \circledast \Gamma(\tau,\tau)$. We demonstrate this technique in Fig.\ref{fig:spectrogram} (b-d), where we have calculated $\Gamma$ once, using only the portion of the spectrum between 25 and 15 fs, where our assumption of slowly changing $A$ is well-justified. Finally, we improve our guess of $A$ by the de-convolution with $\Gamma(\tau,\tau)$.

The reconstruction of $\Gamma(\tau,\tau)$ quickly converges to  the correct average power. It also captures the increasing coherence length towards the head of the pulse. However, the simulated $\Gamma(\tau,\tau)$ shows small coherent satellites near the tail (corresponding to pulse splitting), which are not recovered in this reconstruction. In order to capture these weak features, one could create a spectrogram with higher precision in $k$ and a larger range of streaking fields, but in practice resolving such weak features will be difficult. Further iteration including non-linear optimization of $A$ (as in \cite{keathley_volkov_2016}) can improve the reconstruction of the sharp current spike near $t=0$, but at the cost of significant computational resources.

Once the streaking amplitude is well-calibrated from an average spectrogram, it is possible to estimate single-shot FEL pulse lengths from streaked spectra. To do so accurately, one should set the pump-probe delay such that the streaking field is large and the x-ray spectrum is negligible. Where this is not possible, one can still make an estimate of single-shot parameters, especially if both the parallel and anti-parallel bunches can be gathered and the x-ray spectrum measured downstream (since a comparison of the two gives direct information about the x-ray chirp).

\section{Beam driven quantum dynamics}

The relativistic electron beam is a uniquely broadband pump pulse that is both faster and stronger than comparable ``single-cycle'' pulses (see Figure\,\ref{fig:cartoon}). It can excite electronic transitions, shift electronic energy levels, and drive large-scale nuclear motion. The pulses are fast enough that they can act impulsively\,\cite{jones_ionization_1993}, and yet strong enough to manipulate wavepacket dynamics. 

The wide range of quantum dynamics triggered by the electron beam pump can be tuned by altering the space-charge field's cut-off frequency and field-strength. We consider two cases (Fig.\,\ref{fig:cartoon}b): a 250\,as 12\,kA current spike, and a 250\,fs 1\,kA flat-top electron beam. The current spike contains ~4\,$\upmu$J within its FWHM profile and extends out to 12\,eV at 3.5\,V/\AA\,; while the flat-top beam contains ~11\,$\upmu$J within its FWHM profile and extends out to 1\,THz at 0.15\,V/\AA\,. Both beams are focused to 20\,$\upmu$m (rms) spot-size as a compromise between pumped-volume and field strength. If an ion-microscope is used to spatially resolve the pumped volume \cite{tsatrafyllis_ion_2016}, then the beams could be designed with a tighter focus in order to further increase the field strength. Indeed, at FACET-II a tighter focus and stronger compression are expected to push beam driven fields into the regime of relativistic optics and QED \cite{yakimenko_facet-ii_2019}.

The resulting physics can be directly probed by the intrinsically synchronized XFEL pulse. In the attosecond modality discussed in the previous sections, which builds on previous work at LCLS\,\cite{duris_tunable_2020,li_attosecond_2022}, we imagine using soft-x ray absorption spectroscopy to make a chemically-resolved diagnostic of ultra-fast motion. But it is also possible to measure structural changes directly from x-ray diffraction. In either case, the x-ray probe provides mechanistic insight into the ultrafast dynamics that can't easily be obtained from an optical probe.

Here we discuss three pump-probe scenarios which highlight the flexibility of the electron beam pump source. Firstly, we consider photo-chemical reactions pumped by the EUV spectral component of an attosecond current spike. Secondly, we consider electronic transitions in a large band-gap dielectric where an attosecond current spike causes direct ionization to the conduction band and then a long THz-like pedestal accelerates the free-carriers. Finally, we discuss using a long flattop electron beam in order to drive large-amplitude ion motion in the model battery solid electrolyte  Na~$\beta/\beta^{\prime\prime}$-alumina. For this case, we estimate how soft-x-ray absorption can be used to track large excursions in the mobile Na ion density and thus create a pathway towards a mechanistic understanding of ionic conductivity and its vibrational origins.

\subsection{Ultrafast photochemistry}

Ultrafast photochemistry provides opportunities for improving our understanding in synthetic chemistry and energy storage by delivering energy directly to a target molecule in an otherwise cold system.
The current spike simulated in this article can drive the electronic excitation significantly faster than conventional ultraviolet~(UV) lasers, which allows us to selectively probe ultrafast reaction pathways. 
It is difficult, however, to predict the photochemical reactivity based on the reactant structure through structure-reactivity relationships (in all but the simplest cases). 
Beyond this, non-adiabatic dynamics near conical intersections~(CIs) in excited electronic states involve an interplay of electronic and nuclear degrees of freedom, which cannot be described within the Born-Oppenheimer approximation, and are responsible for much of the excited state behavior in photochemistry. 
By shaping the current profile of our electron beam, we may be able to exert control over the conical interaction and alter the non-adiabatic dynamics \,\cite{arasaki_optical_2010,arasaki_monitoring_2011,richter_sub-laser-cycle_2015,richter_ultrafast_2019} .

\begin{figure}
    \centering
    \includegraphics[scale=0.5]{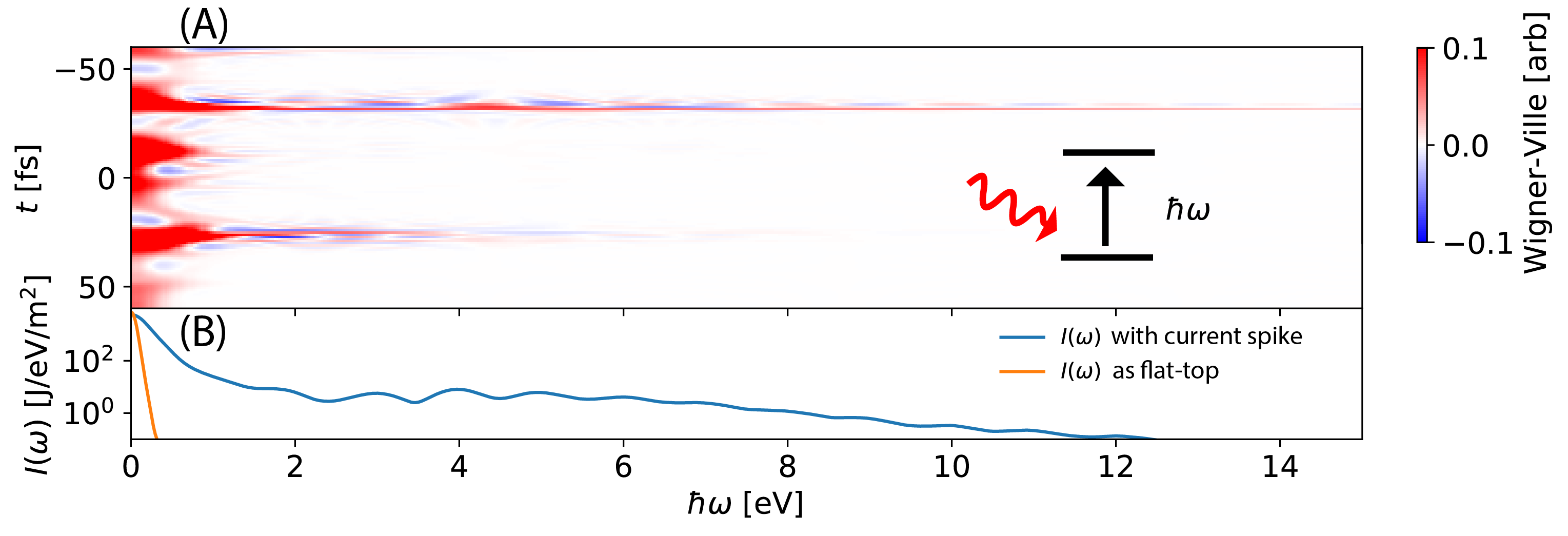}
    \caption{(a) Wigner-Ville distribution of the Coulomb field. (b) Spectrum of the Coulomb field with and without the current spike. With the current spike, the spectrum extends out well into the EUV where it can directly excite many valence transitions. Without the current spike, the spectrum does not extend beyond the THz frequencies. The compression of the beam can be tuned to control the cutoff frequency.}
    \label{fig:wigner_ville}
\end{figure}

The role of a short current spike in driving photo-chemical reactions can be thought of as arising from the EUV-field associated with electromagnetic field of the beam. 
We visualize this effect in Fig.\,\ref{fig:wigner_ville} for the 250\,as spike, shown in Fig.\,\ref{fig:pulse_synthesis} by calculating the time-frequency Wigner-Ville distribution of the electric field. The short spike in the beam leads to a 250\,as window with spectral content out to 12\,eV.  The pedestal accompanying that short spike can cause strong-field ionization if the beam is focused too tightly, but in the perturbative regime it simply causes a Stark-shift of the energy levels. In this case, the ultrafast spike can transition valence electrons into excited states on an impulsive time-scale.

\subsection{Electronic dynamics in a model dielectric}

\begin{figure}
    \centering
    \includegraphics[scale=0.8]{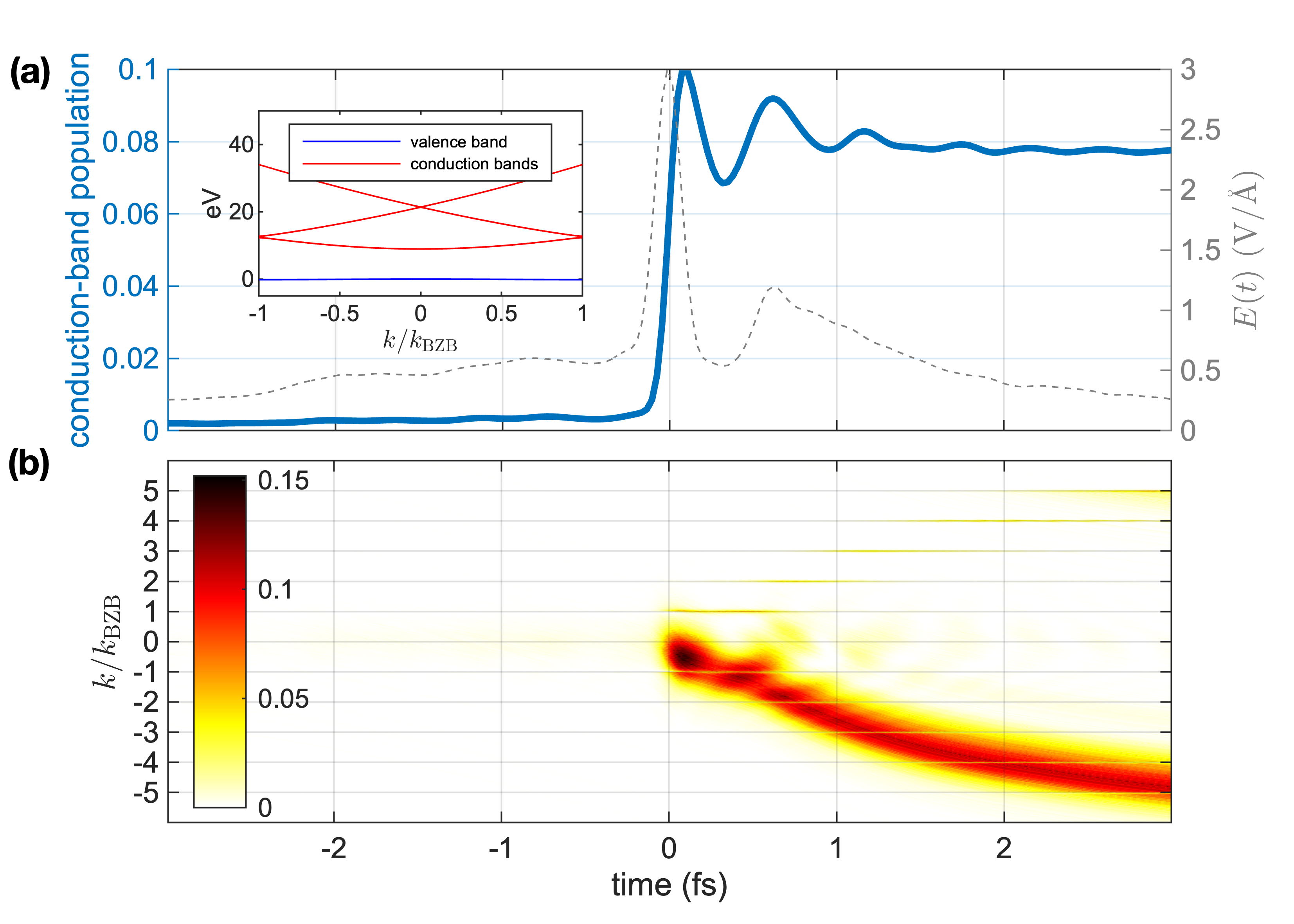}
    \caption{The interaction of the space-charge field (grey dashed curve) with a one-dimensional periodic lattice potential that has a band gap of 9 eV (equivalent to  the band gap of crystalline quartz). (a) From the time evolution of the conduction-band population (blue curve), we see that transitions from valence to conduction bands mainly happen during the main spike of the electric field. The inset shows the band structure within the first Brillouin zone ($k_\mathrm{BZB} = 0.628 \mbox{Å}^{-1}$). (b) Photo-injected electrons are rapidly accelerated by the space-charge field, crossing multiple Brillouin-zone boundaries (BZB), which promotes them to high conduction bands. While panel (a) shows the occupation averaged over crystal momenta, this pseudocolor diagram represents the motion of the electron wave packet in reciprocal space, where the bands are unfolded into the extended-zone scheme, and the colors in this diagram represent the occupations of Bloch states in the first five conduction bands.}
    \label{fig:1Dsolid}
\end{figure}
 
Studies  of  light-matter  interaction  at  the  ultrafast  time-scale  can give unique insight into the optoelectronic  properties  and  device  physics  of  emerging  materials.  Specifically,  the charge-carrier  dynamics  following  light  excitation  have  direct  implications for  optoelectronic  device performance in terms of charge-carrier generations, recombination, and charge-transfer processes. High  intensity fields can be used to drive electrons far from their equilibrium, leading to Bloch oscillations within each sub-band of a solid and processes like high harmonic generation \cite{Yue_22}. The space-charge field opens a new regime for controlling electron dynamics in solids by providing a strong field which can rapidly excite free carriers and drive them to high energies before scattering. Here, as a proof-of-concept, we have performed simulations on the space-charge field interaction within a dielectric. The dielectric is simplified in our simulations with a one-dimensional periodic lattice potential that has a band gap of 9 eV, close to the band gap of crystalline quartz (see Fig.\,\ref{fig:1Dsolid}a)). We solved the time-dependent Schr\"{o}dinger equation in the basis of accelerated Bloch states, as described in \cite{hawkins_effect_2015}. In this model, all electrons initially occupy valence states in a periodic potential, the parameters of which were identical to those in \cite{kruchinin_theory_2013}. The broad spectrum of the space-charge field allows for single-photon transitions across the 9-eV band gap, which are largely confined to the central spike. The spike itself is too short to significantly accelerate the charge carriers that it creates, but the electric field that follows the spike makes electrons acquire kinetic energies on the order of tens of electronvolts. Electrons gain energy by making transitions to higher conduction bands as they cross the Brillouin-zone borders. While Fig.\,\ref{fig:1Dsolid}a) shows the occupation averaged over crystal momenta, Fig.\,\ref{fig:1Dsolid}b) represents the motion of the electron wave packet in reciprocal space. The simulations show that probing the unfolding changes in charge motion upon excitation by the space-charge field with attosecond time resolution and atomic specificity would open up exciting perspectives in attosecond material science, including the development of petahertz optoelectronic switches.

\subsection{Nuclear motion in ionic conductors}

 For frequencies up to several THz, the spectral intensities and peak fields of the space-charge field can be more than $10\times$ those of single-cycle tabletop sources \cite{salen_matter_2019}. This makes it an attractive strong-field source for triggering lattice dynamics that couple to exotic non-equilibrium phases \cite{li_terahertz_2019, de_la_torre_colloquium_2021} or large-amplitude ionic motions that couple to ionic conduction \cite{sood_electrochemical_2021,poletayev_persistence_2021}. Soft x-ray radiation from the same electron beam can then provide chemical resolution to selectively target shifts in the active ions and to probe the time-resolved ion trajectory and associated intermediate states. This would enable a new view of the microscopic processes that underlie how ion motion occurs in rechargeable batteries and other electrochemical systems. \cite{sood_electrochemical_2021}. 

\begin{figure}
    \centering
    \includegraphics[width=5in]{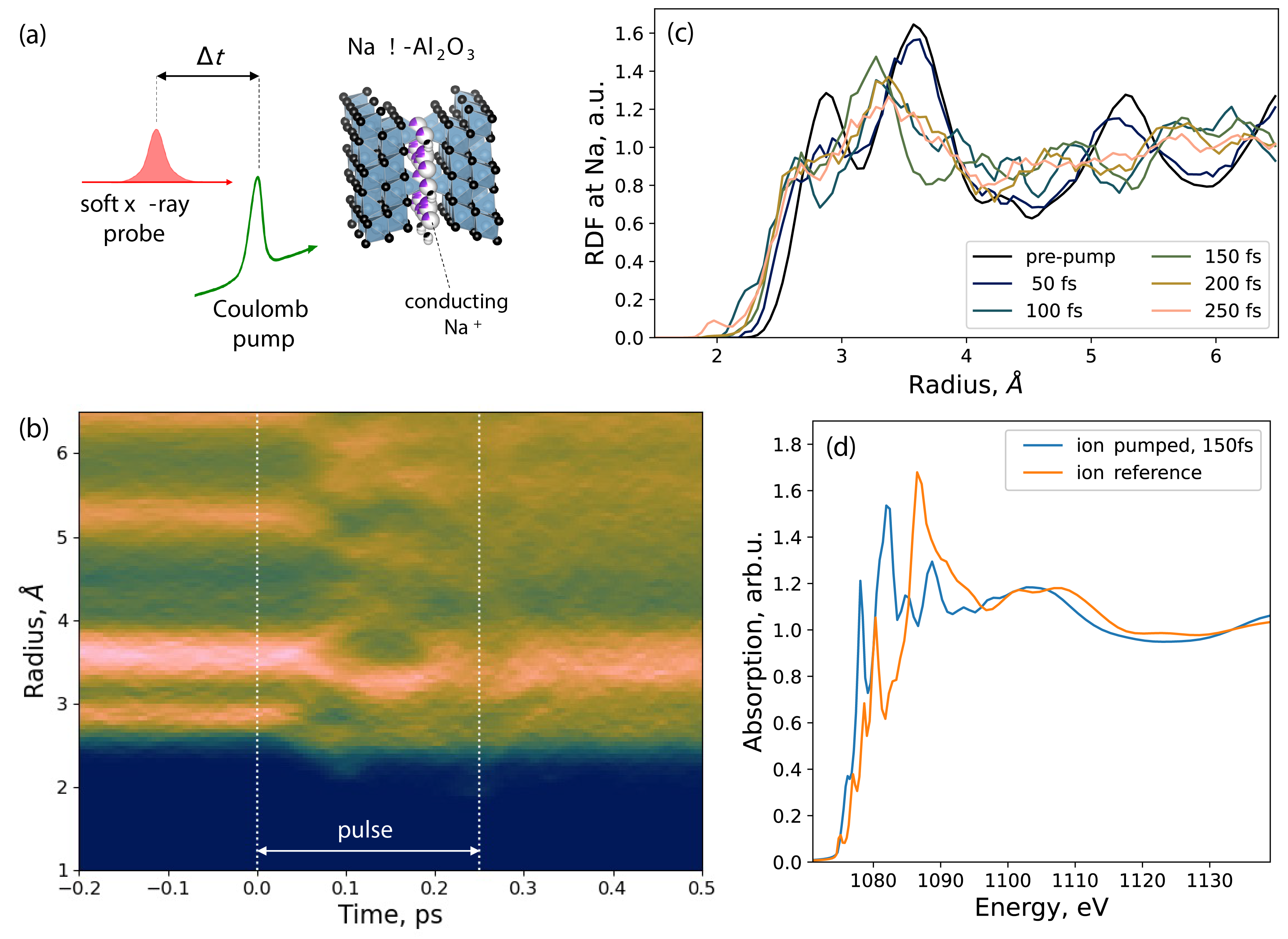}
    \caption{Space charge driven large-amplitude ionic motion of mobile Na$^+$ cations in the model ion conductor Na~$\beta$-alumina. (a) Proposed experiment setup where the terahertz-frequency field acts as a pump of ionic motion followed by a soft x-ray probe from the same FEL pulse. (b,c) Simulated radial distribution function around the Na ions. Beyond 50\,fs the distribution is strongly perturbed by the space charge field. (d) Simulated x-ray absorption spectra of one Na$^+$ ion compared with and without ("reference") the Coulomb-field perturbation. }
    \label{fig:na_beta_alumina}
\end{figure}

Applications include pure ionic conductors such as Na~$\beta/\beta^{\prime\prime}$-aluminas \cite{poletayev_defect-driven_2021,poletayev_persistence_2021}, mixed ion-electron conductors such as the layered-oxide cathodes Li(Ni,Mn,Co)O$_2$ \cite{house_first-cycle_2020,gent_design_2020} or Na$_2$Mn$_3$O$_7$\,\cite{abate_coulombically-stabilized_2021}, quantum paraelectrics such as SrTiO3 \cite{li_terahertz_2019}, and many others. Here, we simulate the pumping of the vibrations of conducting mobile Na$^+$ ions in Na~$\beta$-alumina (Figure~\ref{fig:na_beta_alumina}a). The strongly anharmonic vibrations couple to translations called "hops" and ultimately to long-range ionic conduction, but this coupling remains challenging to both trigger and probe due to fluctuating potential-energy landscapes and the rarity of the hopping events \cite{poletayev_persistence_2021}. The ionic response to a strong-field pump simulated with large-scale molecular dynamics \cite{mishra_ultrafast_2015,mishra_prospects_2018,poletayev_defect-driven_2021,poletayev_persistence_2021} shows a substantially perturbed radial distribution function (RDF) around the mobile Na ions (Figure~\ref{fig:na_beta_alumina}bc) due to their rapid displacements driven by the pump. The x-ray absorption spectrum of a representative ion computed in FEFF\,\cite{kas_advanced_2021} (see apendix) based on the molecular-dynamics trajectory also shows strong changes relative to the same ion in an unperturbed material (Figure~\ref{fig:na_beta_alumina}d), both in the near-edge and EXAFS regions. 

This simplified model suggests that core-level absorption spectroscopy can be used to track the ultrafast ionic motion driven by the strong field of the beam. Furthermore, since the timescales of lattice dynamics and especially metastable states accessed by strong-field excitations can extend much longer than the temporal length of the FEL pulses themselves, additional probes can be subsequently employed for multi-modal characterization of the dynamics triggered by the Coulomb pump.

\section{Conclusions}
In this article we have discussed how to produce and characterize ultra-fast pump-probe fields at an FEL facility. By deliberately seeding coherent microbunches in an electron beam, we can create pairs of ultrashort current spikes. The tail spike generates sub-fs x-ray pulses which then slip ahead to overlap with the head spike resulting in an intrinsically synchronized pump-probe scheme between x-rays and the space-charge field of the beam.

We show that, compared to a conventional laser, the relativistic Coulomb field has unique properties which lend it to novel studies of ultrafast and strong-field dynamics. It can be compressed to attosecond pulse lengths and focused to atomic field strengths, all while supporting frequency content from 0 to 12\,eV. Within this novel parameter space, we find that the beam can be used to explore ultrafast reaction pathways in photo-chemistry, to study opto-electronic proprieties of materials, and to drive large amplitude nuclear motion. The combination of short pulses and large momentum transfer will allow the space charge field to test the limits of material proprieties and improve our understanding of energy transfer on ultrafast time scales.

We support our assertion that an electron beam can be shaped to create both attosecond soft x-ray pulses and a powerful space-charge field by showing start-to-end simulations of the LCLS-II facility. The specificity of our example belies the flexibility of our technique. Indeed, a programmable laser heater can be used to etch complicated shapes into the electron beam current profile\,\cite{cesar_electron_2021}. By synthesizing electric field transients with sub-femtosecond features we can control the potential landscape and thus the dynamics of the resulting excitation\,\cite{arasaki_optical_2010, arasaki_monitoring_2011,richter_sub-laser-cycle_2015,richter_ultrafast_2019}. And with the addition of a dedicated post-lasing compressor we will be able to create high-contrast current spikes capable of cleanly driving the impulsive excitation of valence electrons--a powerful method for wave-packet control which has previously been available only to Rydberg electrons\,\cite{jones_ionization_1993}. The opportunity to not only generate excited states, but also to control them and then probe them with soft x-rays would open a new regime of attosecond physics only possible at state-of-the-art x-ray free electron lasers.

\begin{backmatter}

\bmsection{Funding} 
This work is supported in part by United States~(US) Department of Energy~(DOE) Contract No. DE-AC02-76SF00515.
This work was supported in part by the US DOE, Office of Science, Office of Basic Energy Sciences (BES), Accelerator and Detector Research program, and the DOE-BES Chemical Sciences, Geosciences, and Biosciences Division (CSGB).

A.K. gratefully acknowledges the Computing Center of the Far Eastern Branch of the Russian Academy of Sciences, 680000 Khabarovsk, Russia. The FEFF computations were carried out using the resources of the Center for Shared Use of Scientific Equipment “Center for Processing and Storage of Scientific Data of the Far Eastern Branch of the Russian Academy of Sciences” \cite{sorokin_information_2017}, funded by the Ministry of Science and Higher Education of the Russian Federation under project No. 075-15-2021-663

\bmsection{Acknowledgments} 
We acknowledge many helpful conversations with David Reis concerning the application of strong beam driven fields to solids.

\bmsection{Disclosures} The authors declare no competing interests.

\bmsection{Data availability} Data underlying the results presented in this paper are not publicly available at this time but may be obtained from the authors upon reasonable request.

\end{backmatter}

\bibliography{references.bib,endnotes.bib}

\begin{thebibliography}{10}
\newcommand{\enquote}[1]{``#1''}

\bibitem{bostedt_linac_2016}
C.~Bostedt, S.~Boutet, D.~M. Fritz, Z.~Huang, H.~J. Lee, H.~T. Lemke,
  A.~Robert, W.~F. Schlotter, J.~J. Turner, and G.~J. Williams, \enquote{Linac
  {Coherent} {Light} {Source}: {The} first five years,}
  {\protect\JournalTitle{Reviews of Modern Physics}} \textbf{88}, 015007
  (2016). Publisher: American Physical Society.

\bibitem{lutman_experimental_2013}
A.~A. Lutman, R.~Coffee, Y.~Ding, Z.~Huang, J.~Krzywinski, T.~Maxwell,
  M.~Messerschmidt, and H.-D. Nuhn, \enquote{Experimental {Demonstration} of
  {Femtosecond} {Two}-{Color} {X}-{Ray} {Free}-{Electron} {Lasers},}
  {\protect\JournalTitle{Physical Review Letters}} \textbf{110}, 134801 (2013).
  Publisher: American Physical Society.

\bibitem{marinelli_high-intensity_2015}
A.~Marinelli, D.~Ratner, A.~A. Lutman, J.~Turner, J.~Welch, F.-J. Decker,
  H.~Loos, C.~Behrens, S.~Gilevich, A.~A. Miahnahri, S.~Vetter, T.~J. Maxwell,
  Y.~Ding, R.~Coffee, S.~Wakatsuki, and Z.~Huang, \enquote{High-intensity
  double-pulse {X}-ray free-electron laser,} {\protect\JournalTitle{Nature
  Communications}} \textbf{6}, 1--6 (2015). Number: 1 Publisher: Nature
  Publishing Group.

\bibitem{lutman_fresh-slice_2016}
A.~A. Lutman, T.~J. Maxwell, J.~P. MacArthur, M.~W. Guetg, N.~Berrah, R.~N.
  Coffee, Y.~Ding, Z.~Huang, A.~Marinelli, S.~Moeller, and J.~C.~U. Zemella,
  \enquote{Fresh-slice multicolour {X}-ray free-electron lasers,}
  {\protect\JournalTitle{Nature Photonics}} \textbf{10}, 745--750 (2016).
  Number: 11 Publisher: Nature Publishing Group.

\bibitem{petrovic_transient_2012}
V.~S. Petrović, M.~Siano, J.~L. White, N.~Berrah, C.~Bostedt, J.~D. Bozek,
  D.~Broege, M.~Chalfin, R.~N. Coffee, J.~Cryan, L.~Fang, J.~P. Farrell, L.~J.
  Frasinski, J.~M. Glownia, M.~Gühr, M.~Hoener, D.~M.~P. Holland, J.~Kim,
  J.~P. Marangos, T.~Martinez, B.~K. McFarland, R.~S. Minns, S.~Miyabe,
  S.~Schorb, R.~J. Sension, L.~S. Spector, R.~Squibb, H.~Tao, J.~G. Underwood,
  and P.~H. Bucksbaum, \enquote{Transient {X}-{Ray} {Fragmentation}: {Probing}
  a {Prototypical} {Photoinduced} {Ring} {Opening},}
  {\protect\JournalTitle{Physical Review Letters}} \textbf{108}, 253006 (2012).
  Publisher: American Physical Society.

\bibitem{cryan_auger_2010}
J.~P. Cryan, J.~M. Glownia, J.~Andreasson, A.~Belkacem, N.~Berrah, C.~I. Blaga,
  C.~Bostedt, J.~Bozek, C.~Buth, L.~F. DiMauro, L.~Fang, O.~Gessner, M.~Guehr,
  J.~Hajdu, M.~P. Hertlein, M.~Hoener, O.~Kornilov, J.~P. Marangos, A.~M.
  March, B.~K. McFarland, H.~Merdji, V.~S. Petrović, C.~Raman, D.~Ray,
  D.~Reis, F.~Tarantelli, M.~Trigo, J.~L. White, W.~White, L.~Young, P.~H.
  Bucksbaum, and R.~N. Coffee, \enquote{Auger {Electron} {Angular}
  {Distribution} of {Double} {Core}-{Hole} {States} in the {Molecular}
  {Reference} {Frame},} {\protect\JournalTitle{Physical Review Letters}}
  \textbf{105}, 083004 (2010). Publisher: American Physical Society.

\bibitem{glownia_time-resolved_2010}
J.~M. Glownia, J.~Cryan, J.~Andreasson, A.~Belkacem, N.~Berrah, C.~I. Blaga,
  C.~Bostedt, J.~Bozek, L.~F. DiMauro, L.~Fang, J.~Frisch, O.~Gessner,
  M.~Gühr, J.~Hajdu, M.~P. Hertlein, M.~Hoener, G.~Huang, O.~Kornilov, J.~P.
  Marangos, A.~M. March, B.~K. McFarland, H.~Merdji, V.~S. Petrovic, C.~Raman,
  D.~Ray, D.~A. Reis, M.~Trigo, J.~L. White, W.~White, R.~Wilcox, L.~Young,
  R.~N. Coffee, and P.~H. Bucksbaum, \enquote{Time-resolved pump-probe
  experiments at the {LCLS},} {\protect\JournalTitle{Optics Express}}
  \textbf{18}, 17620--17630 (2010). Publisher: Optical Society of America.

\bibitem{duris_tunable_2020}
J.~Duris, S.~Li, T.~Driver, E.~G. Champenois, J.~P. MacArthur, A.~A. Lutman,
  Z.~Zhang, P.~Rosenberger, J.~W. Aldrich, R.~Coffee, G.~Coslovich, F.-J.
  Decker, J.~M. Glownia, G.~Hartmann, W.~Helml, A.~Kamalov, J.~Knurr,
  J.~Krzywinski, M.-F. Lin, J.~P. Marangos, M.~Nantel, A.~Natan, J.~T.
  O’Neal, N.~Shivaram, P.~Walter, A.~L. Wang, J.~J. Welch, T.~J.~A. Wolf,
  J.~Z. Xu, M.~F. Kling, P.~H. Bucksbaum, A.~Zholents, Z.~Huang, J.~P. Cryan,
  and A.~Marinelli, \enquote{Tunable isolated attosecond {X}-ray pulses with
  gigawatt peak power from a free-electron laser,}
  {\protect\JournalTitle{Nature Photonics}} \textbf{14}, 30--36 (2020). Number:
  1 Publisher: Nature Publishing Group.

\bibitem{li_attosecond_2022}
S.~Li, T.~Driver, P.~Rosenberger, E.~G. Champenois, J.~Duris, A.~Al-Haddad,
  V.~Averbukh, J.~C.~T. Barnard, N.~Berrah, C.~Bostedt, P.~H. Bucksbaum, R.~N.
  Coffee, L.~F. DiMauro, L.~Fang, D.~Garratt, A.~Gatton, Z.~Guo, G.~Hartmann,
  D.~Haxton, W.~Helml, Z.~Huang, A.~C. LaForge, A.~Kamalov, J.~Knurr, M.-F.
  Lin, A.~A. Lutman, J.~P. MacArthur, J.~P. Marangos, M.~Nantel, A.~Natan,
  R.~Obaid, J.~T. O’Neal, N.~H. Shivaram, A.~Schori, P.~Walter, A.~L. Wang,
  T.~J.~A. Wolf, Z.~Zhang, M.~F. Kling, A.~Marinelli, and J.~P. Cryan,
  \enquote{Attosecond coherent electron motion in {Auger}-{Meitner} decay,}
  {\protect\JournalTitle{Science}} \textbf{375}, 285--290 (2022). Publisher:
  American Association for the Advancement of Science.

\bibitem{seltzer_cross_1988}
S.~M. Seltzer, \enquote{Cross {Sections} for {Bremsstrahlung} {Production} and
  {Electron}-{Impact} {Ionization},} in \emph{Monte {Carlo} {Transport} of
  {Electrons} and {Photons},}  T.~M. Jenkins, W.~R. Nelson, and A.~Rindi, eds.
  (Springer US, Boston, MA, 1988), Ettore {Majorana} {International} {Science}
  {Series}, pp. 81--114.

\bibitem{krejcik_commissioning_2003}
P.~Krejcik, \enquote{Commissioning of the {SPPS} {Linac} {Bunch} {Compressor},}
   (2003), p.~3.

\bibitem{clarke_facet_nodate}
C.~I. Clarke, F.~J. Decker, R.~Erikson, C.~Hast, M.~J. Hogan, R.~Iverson, S.~Z.
  Li, Y.~Nosochkov, N.~Phinney, J.~Sheppard, U.~Wienands, M.~Woodley, G.~Yocky,
  M.~Park, A.~Seryi, and W.~Wittmer, \enquote{{FACET}: {THE} {NEW} {USER}
  {FACILITY} {AT} {SLAC},} p.~3.

\bibitem{tudosa_ultimate_2004}
I.~Tudosa, C.~Stamm, A.~B. Kashuba, F.~King, H.~C. Siegmann, J.~Stöhr, G.~Ju,
  B.~Lu, and D.~Weller, \enquote{The ultimate speed of magnetic switching in
  granular recording media,} {\protect\JournalTitle{Nature}} \textbf{428},
  831--833 (2004). Number: 6985 Publisher: Nature Publishing Group.

\bibitem{lu_electronic_2019}
Y.~Lu, A.~Alvarez, C.-H. Kao, J.-S. Bow, S.-Y. Chen, and I.-W. Chen,
  \enquote{An electronic silicon-based memristor with a high switching
  uniformity,} {\protect\JournalTitle{Nature Electronics}} \textbf{2}, 66--74
  (2019). Number: 2 Publisher: Nature Publishing Group.

\bibitem{oshea_conductivity_2019}
B.~D. O’Shea, G.~Andonian, S.~K. Barber, C.~I. Clarke, P.~D. Hoang, M.~J.
  Hogan, B.~Naranjo, O.~B. Williams, V.~Yakimenko, and J.~B. Rosenzweig,
  \enquote{Conductivity {Induced} by {High}-{Field} {Terahertz} {Waves} in
  {Dielectric} {Material},} {\protect\JournalTitle{Physical Review Letters}}
  \textbf{123}, 134801 (2019). Publisher: American Physical Society.

\bibitem{zhang_double_2019}
Z.~Zhang, J.~Duris, J.~P. MacArthur, Z.~Huang, and A.~Marinelli,
  \enquote{Double chirp-taper x-ray free-electron laser for attosecond
  pump-probe experiments,} {\protect\JournalTitle{Physical Review Accelerators
  and Beams}} \textbf{22}, 050701 (2019). Publisher: American Physical Society.

\bibitem{macarthur_phase-stable_2019}
J.~P. MacArthur, J.~Duris, Z.~Zhang, A.~Lutman, A.~Zholents, X.~Xu, Z.~Huang,
  and A.~Marinelli, \enquote{Phase-{Stable} {Self}-{Modulation} of an
  {Electron} {Beam} in a {Magnetic} {Wiggler},} {\protect\JournalTitle{Physical
  Review Letters}} \textbf{123}, 214801 (2019). Publisher: American Physical
  Society.

\bibitem{kang_hard_2017}
H.-S. Kang, C.-K. Min, H.~Heo, C.~Kim, H.~Yang, G.~Kim, I.~Nam, S.~Y. Baek,
  H.-J. Choi, G.~Mun, B.~R. Park, Y.~J. Suh, D.~C. Shin, J.~Hu, J.~Hong,
  S.~Jung, S.-H. Kim, K.~Kim, D.~Na, S.~S. Park, Y.~J. Park, J.-H. Han, Y.~G.
  Jung, S.~H. Jeong, H.~G. Lee, S.~Lee, S.~Lee, W.-W. Lee, B.~Oh, H.~S. Suh,
  Y.~W. Parc, S.-J. Park, M.~H. Kim, N.-S. Jung, Y.-C. Kim, M.-S. Lee, B.-H.
  Lee, C.-W. Sung, I.-S. Mok, J.-M. Yang, C.-S. Lee, H.~Shin, J.~H. Kim,
  Y.~Kim, J.~H. Lee, S.-Y. Park, J.~Kim, J.~Park, I.~Eom, S.~Rah, S.~Kim, K.~H.
  Nam, J.~Park, J.~Park, S.~Kim, S.~Kwon, S.~H. Park, K.~S. Kim, H.~Hyun, S.~N.
  Kim, S.~Kim, S.-m. Hwang, M.~J. Kim, C.-y. Lim, C.-J. Yu, B.-S. Kim, T.-H.
  Kang, K.-W. Kim, S.-H. Kim, H.-S. Lee, H.-S. Lee, K.-H. Park, T.-Y. Koo,
  D.-E. Kim, and I.~S. Ko, \enquote{Hard {X}-ray free-electron laser with
  femtosecond-scale timing jitter,} {\protect\JournalTitle{Nature Photonics}}
  \textbf{11}, 708--713 (2017). Number: 11 Publisher: Nature Publishing Group.

\bibitem{hartmann_sub-femtosecond_2014}
N.~Hartmann, W.~Helml, A.~Galler, M.~R. Bionta, J.~Grünert, S.~L. Molodtsov,
  K.~R. Ferguson, S.~Schorb, M.~L. Swiggers, S.~Carron, C.~Bostedt, J.-C.
  Castagna, J.~Bozek, J.~M. Glownia, D.~J. Kane, A.~R. Fry, W.~E. White, C.~P.
  Hauri, T.~Feurer, and R.~N. Coffee, \enquote{Sub-femtosecond precision
  measurement of relative {X}-ray arrival time for free-electron lasers,}
  {\protect\JournalTitle{Nature Photonics}} \textbf{8}, 706--709 (2014).
  Number: 9 Publisher: Nature Publishing Group.

\bibitem{woodley_lcls_nodate}
M.~Woodley, \enquote{{LCLS} {Lattice} {Description},} .

\bibitem{zhang_high-power_2020}
Z.~Zhang, A.~S. Fisher, M.~C. Hoffmann, B.~Jacobson, P.~S. Kirchmann, W.-S.
  Lee, A.~Lindenberg, A.~Marinelli, E.~Nanni, R.~Schoenlein, M.~Qian,
  S.~Sasaki, J.~Xu, and Z.~Huang, \enquote{A high-power, high-repetition-rate
  {THz} source for pump–probe experiments at {Linac} {Coherent} {Light}
  {Source} {II},} {\protect\JournalTitle{Journal of Synchrotron Radiation}}
  \textbf{27}, 890--901 (2020). Number: 4 Publisher: International Union of
  Crystallography.

\bibitem{koevener_thz_2016}
T.~Koevener, \enquote{{THz} spectrometer calibration at {FELIX},} Ph.D. thesis,
  Universität Hamburg, Masterarbeit, 2016 (2016). Number: PUBDB-2016-06662.

\bibitem{ozerov_thz_2017}
M.~Ozerov, B.~Bern\'{a}th, D.~Kamenskyi, B.~Redlich, A.~F.~G. van~der Meer,
  P.~C.~M. Christianen, H.~Engelkamp, and J.~C. Maan, \enquote{A {THz}
  spectrometer combining the free electron laser {FLARE} with 33 {T} magnetic
  fields,} {\protect\JournalTitle{Applied Physics Letters}} \textbf{110},
  094106 (2017). Publisher: American Institute of Physics.

\bibitem{bakker_intensity-resolved_2010}
J.~M. Bakker, V.~J.~F. Lapoutre, B.~Redlich, J.~Oomens, B.~G. Sartakov,
  A.~Fielicke, G.~von Helden, G.~Meijer, and A.~F.~G. van~der Meer,
  \enquote{Intensity-resolved {IR} multiple photon ionization and fragmentation
  of {C60},} {\protect\JournalTitle{The Journal of Chemical Physics}}
  \textbf{132}, 074305 (2010). Publisher: American Institute of Physics.

\bibitem{li_heting__2021}
L.~Heting, H.~Zhigang, W.~Fangfang, T.~Leilei, Z.~Zhouyu, Z.~Tong, H.~Tianlong,
  X.~Ke, Z.~Haiyan, W.~Wei, L.~Ping, Z.~Zeran, S.~Lei, L.~Gongfa, X.~Hongliang,
  H.~Xiaoye, J.~Shiping, P.~Xiangtao, G.~Chen, J.~Qika, B.~Jun, Z.~Shancai, and
  W.~Lin, \enquote{Hefei infrared free electron laser facility,}
  {\protect\JournalTitle{Chinese Journal of Lasers}} \textbf{48}, 1700001
  (2021).

\bibitem{ortega_extension_2006}
J.~M. Ortega, F.~Glotin, and R.~Prazeres, \enquote{Extension in far-infrared of
  the {CLIO} free-electron laser,} {\protect\JournalTitle{Infrared Physics \&
  Technology}} \textbf{49}, 133--138 (2006).

\bibitem{kawase_extremely_2020}
K.~Kawase, M.~Nagai, K.~Furukawa, M.~Fujimoto, R.~Kato, Y.~Honda, and
  G.~Isoyama, \enquote{Extremely high-intensity operation of a {THz}
  free-electron laser using an electron beam with a higher bunch charge,}
  {\protect\JournalTitle{Nuclear Instruments and Methods in Physics Research
  Section A: Accelerators, Spectrometers, Detectors and Associated Equipment}}
  \textbf{960}, 163582 (2020).

\bibitem{shalaby_demonstration_2015}
M.~Shalaby and C.~P. Hauri, \enquote{Demonstration of a low-frequency
  three-dimensional terahertz bullet with extreme brightness,}
  {\protect\JournalTitle{Nature Communications}} \textbf{6}, 5976 (2015).

\bibitem{vicario_generation_2014}
C.~Vicario, A.~V. Ovchinnikov, S.~I. Ashitkov, M.~B. Agranat, V.~E. Fortov, and
  C.~P. Hauri, \enquote{Generation of 0.9-{mJ} {THz} pulses in {DSTMS} pumped
  by a {Cr}:{Mg}$_{\textrm{2}}${SiO}$_{\textrm{4}}$ laser,}
  {\protect\JournalTitle{Optics Letters}} \textbf{39}, 6632--6635 (2014).
  Publisher: Optica Publishing Group.

\bibitem{sell_phase-locked_2008}
A.~Sell, A.~Leitenstorfer, and R.~Huber, \enquote{Phase-locked generation and
  field-resolved detection of widely tunable terahertz pulses with amplitudes
  exceeding 100 {MV}/cm,} {\protect\JournalTitle{Optics Letters}} \textbf{33},
  2767--2769 (2008).

\bibitem{wu_highly_2018}
X.-j. Wu, J.-l. Ma, B.-l. Zhang, S.-s. Chai, Z.-j. Fang, C.-Y. Xia, D.-y. Kong,
  J.-g. Wang, H.~Liu, C.-Q. Zhu, X.~Wang, C.-J. Ruan, and Y.-T. Li,
  \enquote{Highly efficient generation of 0.2 {mJ} terahertz pulses in lithium
  niobate at room temperature with sub-50 fs chirped {Ti}:sapphire laser
  pulses,} {\protect\JournalTitle{Optics Express}} \textbf{26}, 7107--7116
  (2018).

\bibitem{hauri_strong-field_2011}
C.~P. Hauri, C.~Ruchert, C.~Vicario, and F.~Ardana, \enquote{Strong-field
  single-cycle {THz} pulses generated in an organic crystal,}
  {\protect\JournalTitle{Applied Physics Letters}} \textbf{99}, 161116 (2011).

\bibitem{fulop_efficient_2014}
J.~A. Fülöp, Z.~Ollmann, C.~Lombosi, C.~Skrobol, S.~Klingebiel, L.~Pálfalvi,
  F.~Krausz, S.~Karsch, and J.~Hebling, \enquote{Efficient generation of {THz}
  pulses with 0.4 {mJ} energy,} {\protect\JournalTitle{Optics Express}}
  \textbf{22}, 20155--20163 (2014).

\bibitem{oh_intense_2013}
T.~I. Oh, Y.~S. You, N.~Jhajj, E.~W. Rosenthal, H.~M. Milchberg, and K.~Y. Kim,
  \enquote{Intense terahertz generation in two-color laser filamentation:
  energy scaling with terawatt laser systems,} {\protect\JournalTitle{New
  Journal of Physics}} \textbf{15}, 075002 (2013).

\bibitem{fulop_highly_2016}
J.~A. Fülöp, G.~Polónyi, B.~Monoszlai, G.~Andriukaitis, T.~Balciunas,
  A.~Pugzlys, G.~Arthur, A.~Baltuska, and J.~Hebling, \enquote{Highly efficient
  scalable monolithic semiconductor terahertz pulse source,}
  {\protect\JournalTitle{Optica}} \textbf{3}, 1075--1078 (2016).

\bibitem{dey_highly_2017}
I.~Dey, K.~Jana, V.~Y. Fedorov, A.~D. Koulouklidis, A.~Mondal, M.~Shaikh,
  D.~Sarkar, A.~D. Lad, S.~Tzortzakis, A.~Couairon, and G.~R. Kumar,
  \enquote{Highly efficient broadband terahertz generation from ultrashort
  laser filamentation in liquids,} {\protect\JournalTitle{Nature
  Communications}} \textbf{8}, 1184 (2017).

\bibitem{yoon_realization_2021}
J.~W. Yoon, J.~W. Yoon, Y.~G. Kim, Y.~G. Kim, I.~W. Choi, I.~W. Choi, J.~H.
  Sung, J.~H. Sung, H.~W. Lee, S.~K. Lee, S.~K. Lee, S.~K. Lee, C.~H. Nam,
  C.~H. Nam, and C.~H. Nam, \enquote{Realization of laser intensity over
  10$^{\textrm{23}}$ {W}/cm$^{\textrm{2}}$,} {\protect\JournalTitle{Optica}}
  \textbf{8}, 630--635 (2021).

\bibitem{shaw_ultrahigh-brightness_1993}
M.~J. Shaw, G.~Bialolenker, G.~J. Hirst, C.~J. Hooker, M.~H. Key, A.~K. Kidd,
  J.~M.~D. Lister, K.~E. Hill, G.~H.~C. New, and D.~C. Wilson,
  \enquote{Ultrahigh-brightness laser beams with low prepulse obtained by
  stimulated {Raman} scattering,} {\protect\JournalTitle{Optics Letters}}
  \textbf{18}, 1320--1322 (1993). Publisher: Optica Publishing Group.

\bibitem{mizoguchi_100-fs_1992}
M.~Mizoguchi, K.~Kondo, and S.~Watanabe, \enquote{100-fs, 10-{Hz}, terawatt
  {KrF} laser,} {\protect\JournalTitle{JOSA B}} \textbf{9}, 560--564 (1992).

\bibitem{liang_high-energy_2017}
H.~Liang, P.~Krogen, Z.~Wang, H.~Park, T.~Kroh, K.~Zawilski, P.~Schunemann,
  J.~Moses, L.~F. DiMauro, F.~X. Kärtner, and K.-H. Hong, \enquote{High-energy
  mid-infrared sub-cycle pulse synthesis from a parametric amplifier,}
  {\protect\JournalTitle{Nature Communications}} \textbf{8}, 141 (2017).

\bibitem{junginger_single-cycle_2010}
F.~Junginger, A.~Sell, O.~Schubert, B.~Mayer, D.~Brida, M.~Marangoni,
  G.~Cerullo, A.~Leitenstorfer, and R.~Huber, \enquote{Single-cycle
  multiterahertz transients with peak fields above 10 {MV}/cm,}
  {\protect\JournalTitle{Optics Letters}} \textbf{35}, 2645--2647 (2010).
  Publisher: Optica Publishing Group.

\bibitem{deng_carrier-envelope-phase-stable_2012}
Y.~Deng, A.~Schwarz, H.~Fattahi, M.~Ueffing, X.~Gu, M.~Ossiander, T.~Metzger,
  V.~Pervak, H.~Ishizuki, T.~Taira, T.~Kobayashi, G.~Marcus, F.~Krausz,
  R.~Kienberger, and N.~Karpowicz, \enquote{Carrier-envelope-phase-stable,
  1.{2mJ}, 1.5 cycle laser pulses at 2.1um,} {\protect\JournalTitle{Optics
  Letters}} \textbf{37}, 4973--4975 (2012).

\bibitem{andriukaitis_90_2011}
G.~Andriukaitis, T.~Balčiūnas, S.~Ališauskas, A.~Pugžlys, A.~Baltuška,
  T.~Popmintchev, M.-C. Chen, M.~M. Murnane, and H.~C. Kapteyn, \enquote{90
  {GW} peak power few-cycle mid-infrared pulses from an optical parametric
  amplifier,} {\protect\JournalTitle{Optics Letters}} \textbf{36}, 2755--2757
  (2011). Publisher: Optica Publishing Group.

\bibitem{grafenstein_multi-millijoule_2020}
L.~v. Grafenstein, M.~Bock, D.~Ueberschaer, E.~Escoto, A.~Koç, K.~Zawilski,
  P.~Schunemann, U.~Griebner, and T.~Elsaesser, \enquote{Multi-millijoule,
  few-cycle 5 µm {OPCPA} at 1 {kHz} repetition rate,}
  {\protect\JournalTitle{Optics Letters}} \textbf{45}, 5998--6001 (2020).
  Publisher: Optica Publishing Group.

\bibitem{elu_table-top_2019}
U.~Elu, T.~Steinle, D.~Sánchez, L.~Maidment, K.~Zawilski, P.~Schunemann, U.~D.
  Zeitner, C.~Simon-Boisson, and J.~Biegert, \enquote{Table-top high-energy
  7\&\#x2009;\&\#x2009;\&\#{x03BC};m {OPCPA} and 260\&\#x2009;\&\#x2009;{mJ}
  {Ho}:{YLF} pump laser,} {\protect\JournalTitle{Optics Letters}} \textbf{44},
  3194--3197 (2019). Publisher: Optica Publishing Group.

\bibitem{travers_high-energy_2019}
J.~C. Travers, T.~F. Grigorova, C.~Brahms, and F.~Belli, \enquote{High-energy
  pulse self-compression and ultraviolet generation through soliton dynamics in
  hollow capillary fibres,} {\protect\JournalTitle{Nature Photonics}}
  \textbf{13}, 547--554 (2019).

\bibitem{ouille_relativistic-intensity_2020}
M.~Ouillé, A.~Vernier, F.~Böhle, M.~Bocoum, A.~Jullien, M.~Lozano, J.-P.
  Rousseau, Z.~Cheng, D.~Gustas, A.~Blumenstein, P.~Simon, S.~Haessler,
  J.~Faure, T.~Nagy, and R.~Lopez-Martens, \enquote{Relativistic-intensity
  near-single-cycle light waveforms at {kHz} repetition rate,}
  {\protect\JournalTitle{Light: Science \& Applications}} \textbf{9}, 47
  (2020). Number: 1 Publisher: Nature Publishing Group.

\bibitem{rivas_next_2017}
D.~E. Rivas, A.~Borot, D.~E. Cardenas, G.~Marcus, X.~Gu, D.~Herrmann, J.~Xu,
  J.~Tan, D.~Kormin, G.~Ma, W.~Dallari, G.~D. Tsakiris, I.~B. Földes, S.-w.
  Chou, M.~Weidman, B.~Bergues, T.~Wittmann, H.~Schröder, P.~Tzallas,
  D.~Charalambidis, O.~Razskazovskaya, V.~Pervak, F.~Krausz, and L.~Veisz,
  \enquote{Next {Generation} {Driver} for {Attosecond} and {Laser}-plasma
  {Physics},} {\protect\JournalTitle{Scientific Reports}} \textbf{7}, 5224
  (2017). Number: 1 Publisher: Nature Publishing Group.

\bibitem{daranciang_single-cycle_2011}
D.~Daranciang, J.~Goodfellow, M.~Fuchs, H.~Wen, S.~Ghimire, D.~A. Reis,
  H.~Loos, A.~S. Fisher, and A.~M. Lindenberg, \enquote{Single-cycle terahertz
  pulses with 0.2{V}/\aa field amplitudes via coherent transition radiation,}
  {\protect\JournalTitle{Applied Physics Letters}} \textbf{99}, 141117 (2011).
  Publisher: American Institute of Physics.

\bibitem{wu_intense_2013}
Z.~Wu, A.~S. Fisher, J.~Goodfellow, M.~Fuchs, D.~Daranciang, M.~Hogan, H.~Loos,
  and A.~Lindenberg, \enquote{Intense terahertz pulses from {SLAC} electron
  beams using coherent transition radiation,} {\protect\JournalTitle{Review of
  Scientific Instruments}} \textbf{84}, 022701 (2013). Publisher: American
  Institute of Physics.

\bibitem{fulop_laser-driven_2020}
J.~A. Fülöp, S.~Tzortzakis, and T.~Kampfrath, \enquote{Laser-{Driven}
  {Strong}-{Field} {Terahertz} {Sources},} {\protect\JournalTitle{Advanced
  Optical Materials}} \textbf{8}, 1900681 (2020). \_eprint:
  https://onlinelibrary.wiley.com/doi/pdf/10.1002/adom.201900681.

\bibitem{hzdr_thz_2022}
HZDR, \enquote{{THz} {FEL} {Sources},}  (2022).

\bibitem{herziger_33_2007}
S.~Szatmári, G.~Marowsky, and P.~Simon, \enquote{3.3 {Femtosecond} excimer
  lasers and their applications,} in \emph{Laser {Systems}, {Part} 1,}  vol.~11
  G.~Herziger, H.~Weber, and R.~Poprawe, eds. (Springer Berlin Heidelberg,
  Berlin, Heidelberg, 2007), pp. 215--253.

\bibitem{focusing_footnote}
In the cases where we did not extract focal-spot data [22-23,25-27,39,42-44] we
  assume a transverse size (rms of intensity distribution) of $5\lambda/\pi$.

\bibitem{cesar_electron_2021}
D.~Cesar, A.~Anakru, S.~Carbajo, J.~Duris, P.~Franz, S.~Li, N.~Sudar, Z.~Zhang,
  and A.~Marinelli, \enquote{Electron beam shaping via laser heater temporal
  shaping,} {\protect\JournalTitle{Physical Review Accelerators and Beams}}
  \textbf{24}, 110703 (2021). Publisher: American Physical Society.

\bibitem{qiang_start--end_2017}
J.~Qiang, Y.~Ding, P.~Emma, Z.~Huang, D.~Ratner, T.~O. Raubenheimer,
  M.~Venturini, and F.~Zhou, \enquote{Start-to-end simulation of the shot-noise
  driven microbunching instability experiment at the {Linac} {Coherent} {Light}
  {Source},} {\protect\JournalTitle{Physical Review Accelerators and Beams}}
  \textbf{20}, 054402 (2017). Publisher: American Physical Society.

\bibitem{wang_benchmark_2015}
L.~Wang, P.~Emma, J.~Qiang, and T.~Raubenheimer, \enquote{Benchmark of
  {ELEGANT} and {IMPACT},} {\protect\JournalTitle{Proceedings of the 37th
  International Free Electron Laser Conference}} \textbf{FEL2015}, 3 pages,
  0.946 MB (2015). Artwork Size: 3 pages, 0.946 MB ISBN: 9783954501342 Medium:
  PDF Publisher: JACoW, Geneva, Switzerland.

\bibitem{borland_elegant_2000}
M.~Borland, \enquote{{ELEGANT}: {A} flexible {SDDS}-compliant code for
  accelerator simulation,} Tech. Rep. LS-287, Argonne National Lab., IL (US)
  (2000).

\bibitem{reiche_genesis_1999}
S.~Reiche, \enquote{{GENESIS} 1.3: a fully {3D} time-dependent {FEL} simulation
  code,} {\protect\JournalTitle{Nuclear Instruments and Methods in Physics
  Research Section A: Accelerators, Spectrometers, Detectors and Associated
  Equipment}} \textbf{429}, 243--248 (1999).

\bibitem{robicheaux_coherent_2000}
F.~Robicheaux and L.~D. Noordam, \enquote{Coherent {Scattering} with {Pulsed}
  {Matter} {Beams},} {\protect\JournalTitle{Physical Review Letters}}
  \textbf{84}, 3735--3739 (2000). Publisher: American Physical Society.

\bibitem{van_der_geer_general_nodate}
S.~B. van~der Geer and M.~J. de~Loos, \enquote{General particle tracer,} .

\bibitem{orfanos_attosecond_2019}
I.~Orfanos, I.~Makos, I.~Liontos, E.~Skantzakis, B.~Förg, D.~Charalambidis,
  and P.~Tzallas, \enquote{Attosecond pulse metrology,}
  {\protect\JournalTitle{APL Photonics}} \textbf{4}, 080901 (2019). Publisher:
  American Institute of Physics.

\bibitem{itatani_attosecond_2002}
J.~Itatani, F.~Quéré, G.~L. Yudin, M.~Y. Ivanov, F.~Krausz, and P.~B. Corkum,
  \enquote{Attosecond {Streak} {Camera},} {\protect\JournalTitle{Physical
  Review Letters}} \textbf{88}, 173903 (2002). Publisher: American Physical
  Society.

\bibitem{drescher_x-ray_2001}
M.~Drescher, M.~Hentschel, R.~Kienberger, G.~Tempea, C.~Spielmann, G.~A.
  Reider, P.~B. Corkum, and F.~Krausz, \enquote{X-ray {Pulses} {Approaching}
  the {Attosecond} {Frontier},} {\protect\JournalTitle{Science}} \textbf{291},
  1923--1927 (2001). Publisher: American Association for the Advancement of
  Science Section: Research Article.

\bibitem{mairesse_frequency-resolved_2005}
Y.~Mairesse and F.~Quéré, \enquote{Frequency-resolved optical gating for
  complete reconstruction of attosecond bursts,}
  {\protect\JournalTitle{Physical Review A}} \textbf{71}, 011401 (2005).
  Publisher: American Physical Society.

\bibitem{gagnon_accurate_2008}
J.~Gagnon, E.~Goulielmakis, and V.~Yakovlev, \enquote{The accurate {FROG}
  characterization of attosecond pulses from streaking measurements,}
  {\protect\JournalTitle{Applied Physics B}} \textbf{92}, 25--32 (2008).

\bibitem{chini_characterizing_2010}
M.~Chini, S.~Gilbertson, S.~D. Khan, and Z.~Chang, \enquote{Characterizing
  ultrabroadband attosecond lasers,} {\protect\JournalTitle{Optics Express}}
  \textbf{18}, 13006--13016 (2010). Publisher: Optica Publishing Group.

\bibitem{keathley_volkov_2016}
P.~D. Keathley, S.~Bhardwaj, J.~Moses, G.~Laurent, and F.~X. Kärtner,
  \enquote{Volkov transform generalized projection algorithm for attosecond
  pulse characterization,} {\protect\JournalTitle{New Journal of Physics}}
  \textbf{18}, 073009 (2016). Publisher: IOP Publishing.

\bibitem{haynes_clocking_2021}
D.~C. Haynes, M.~Wurzer, A.~Schletter, A.~Al-Haddad, C.~Blaga, C.~Bostedt,
  J.~Bozek, H.~Bromberger, M.~Bucher, A.~Camper, S.~Carron, R.~Coffee, J.~T.
  Costello, L.~F. DiMauro, Y.~Ding, K.~Ferguson, I.~Grguraš, W.~Helml, M.~C.
  Hoffmann, M.~Ilchen, S.~Jalas, N.~M. Kabachnik, A.~K. Kazansky,
  R.~Kienberger, A.~R. Maier, T.~Maxwell, T.~Mazza, M.~Meyer, H.~Park,
  J.~Robinson, C.~Roedig, H.~Schlarb, R.~Singla, F.~Tellkamp, P.~A. Walker,
  K.~Zhang, G.~Doumy, C.~Behrens, and A.~L. Cavalieri, \enquote{Clocking
  {Auger} electrons,} {\protect\JournalTitle{Nature Physics}} \textbf{17},
  512--518 (2021). Number: 4 Publisher: Nature Publishing Group.

\bibitem{hartmann_attosecond_2018}
N.~Hartmann, G.~Hartmann, R.~Heider, M.~S. Wagner, M.~Ilchen, J.~Buck, A.~O.
  Lindahl, C.~Benko, J.~Grünert, J.~Krzywinski, J.~Liu, A.~A. Lutman,
  A.~Marinelli, T.~Maxwell, A.~A. Miahnahri, S.~P. Moeller, M.~Planas,
  J.~Robinson, A.~K. Kazansky, N.~M. Kabachnik, J.~Viefhaus, T.~Feurer,
  R.~Kienberger, R.~N. Coffee, and W.~Helml, \enquote{Attosecond time–energy
  structure of {X}-ray free-electron laser pulses,}
  {\protect\JournalTitle{Nature Photonics}} \textbf{12}, 215--220 (2018).
  Number: 4 Publisher: Nature Publishing Group.

\bibitem{heider_megahertz-compatible_2019}
R.~Heider, M.~S. Wagner, N.~Hartmann, M.~Ilchen, J.~Buck, G.~Hartmann,
  V.~Shirvanyan, A.~O. Lindahl, J.~Grünert, J.~Krzywinski, J.~Liu,
  M.~Ossiander, A.~A. Lutman, T.~Maxwell, A.~A. Miahnahri, S.~P. Moeller,
  M.~Planas, J.~Robinson, J.~Viefhaus, T.~Feurer, R.~Kienberger, R.~N. Coffee,
  and W.~Helml, \enquote{Megahertz-compatible angular streaking with
  few-femtosecond resolution at x-ray free-electron lasers,}
  {\protect\JournalTitle{Physical Review A}} \textbf{100}, 053420 (2019).
  Publisher: American Physical Society.

\bibitem{li_characterizing_2018}
S.~Li, Z.~Guo, R.~N. Coffee, K.~Hegazy, Z.~Huang, A.~Natan, T.~Osipov, D.~Ray,
  A.~Marinelli, and J.~P. Cryan, \enquote{Characterizing isolated attosecond
  pulses with angular streaking,} {\protect\JournalTitle{Optics Express}}
  \textbf{26}, 4531--4547 (2018). Publisher: Optical Society of America.

\bibitem{jones_ionization_1993}
R.~R. Jones, D.~You, and P.~H. Bucksbaum, \enquote{Ionization of {Rydberg}
  atoms by subpicosecond half-cycle electromagnetic pulses,}
  {\protect\JournalTitle{Physical Review Letters}} \textbf{70}, 1236--1239
  (1993). Publisher: American Physical Society.

\bibitem{tsatrafyllis_ion_2016}
N.~Tsatrafyllis, B.~Bergues, H.~Schröder, L.~Veisz, E.~Skantzakis, D.~Gray,
  B.~Bodi, S.~Kuhn, G.~D. Tsakiris, D.~Charalambidis, and P.~Tzallas,
  \enquote{The ion microscope as a tool for quantitative measurements in the
  extreme ultraviolet,} {\protect\JournalTitle{Scientific Reports}} \textbf{6},
  21556 (2016). Number: 1 Publisher: Nature Publishing Group.

\bibitem{yakimenko_facet-ii_2019}
V.~Yakimenko, L.~Alsberg, E.~Bong, G.~Bouchard, C.~Clarke, C.~Emma, S.~Green,
  C.~Hast, M.~J. Hogan, J.~Seabury, N.~Lipkowitz, B.~O’Shea, D.~Storey,
  G.~White, and G.~Yocky, \enquote{{FACET}-{II} facility for advanced
  accelerator experimental tests,} {\protect\JournalTitle{Physical Review
  Accelerators and Beams}} \textbf{22}, 101301 (2019).

\bibitem{arasaki_optical_2010}
Y.~Arasaki and K.~Takatsuka, \enquote{Optical conversion of conical
  intersection to avoided crossing,} {\protect\JournalTitle{Physical Chemistry
  Chemical Physics}} \textbf{12}, 1239--1242 (2010). Publisher: The Royal
  Society of Chemistry.

\bibitem{arasaki_monitoring_2011}
Y.~Arasaki, K.~Wang, V.~McKoy, and K.~Takatsuka, \enquote{Monitoring the effect
  of a control pulse on a conical intersection by time-resolved photoelectron
  spectroscopy,} {\protect\JournalTitle{Physical Chemistry Chemical Physics}}
  \textbf{13}, 8681--8689 (2011). Publisher: The Royal Society of Chemistry.

\bibitem{richter_sub-laser-cycle_2015}
M.~Richter, F.~Bouakline, J.~González-Vázquez, L.~Martínez-Fernández,
  I.~Corral, S.~Patchkovskii, F.~Morales, M.~Ivanov, F.~Martín, and
  O.~Smirnova, \enquote{Sub-laser-cycle control of coupled electron–nuclear
  dynamics at a conical intersection,} {\protect\JournalTitle{New Journal of
  Physics}} \textbf{17}, 113023 (2015). Publisher: IOP Publishing.

\bibitem{richter_ultrafast_2019}
M.~Richter, J.~González-Vázquez, Z.~Mašín, D.~S. Brambila, A.~G. Harvey,
  F.~Morales, and F.~Martín, \enquote{Ultrafast imaging of laser-controlled
  non-adiabatic dynamics in {NO2} from time-resolved photoelectron emission,}
  {\protect\JournalTitle{Physical Chemistry Chemical Physics}} \textbf{21},
  10038--10051 (2019). Publisher: The Royal Society of Chemistry.

\bibitem{Yue_22}
L.~Yue and M.~B. Gaarde, \enquote{Introduction to theory of high-harmonic
  generation in solids: tutorial,} {\protect\JournalTitle{J. Opt. Soc. Am. B}}
  \textbf{39}, 535--555 (2022).

\bibitem{hawkins_effect_2015}
P.~G. Hawkins, M.~Y. Ivanov, and V.~S. Yakovlev, \enquote{Effect of multiple
  conduction bands on high-harmonic emission from dielectrics,}
  {\protect\JournalTitle{Physical Review A}} \textbf{91}, 013405 (2015).
  Publisher: American Physical Society.

\bibitem{kruchinin_theory_2013}
S.~Y. Kruchinin, M.~Korbman, and V.~S. Yakovlev, \enquote{Theory of
  strong-field injection and control of photocurrent in dielectrics and wide
  band gap semiconductors,} {\protect\JournalTitle{Physical Review B}}
  \textbf{87}, 115201 (2013). Publisher: American Physical Society.

\bibitem{salen_matter_2019}
P.~Salén, M.~Basini, S.~Bonetti, J.~Hebling, M.~Krasilnikov, A.~Y. Nikitin,
  G.~Shamuilov, Z.~Tibai, V.~Zhaunerchyk, and V.~Goryashko, \enquote{Matter
  manipulation with extreme terahertz light: {Progress} in the enabling {THz}
  technology,} {\protect\JournalTitle{Physics Reports}} \textbf{836-837}, 1--74
  (2019).

\bibitem{li_terahertz_2019}
X.~Li, T.~Qiu, J.~Zhang, E.~Baldini, J.~Lu, A.~M. Rappe, and K.~A. Nelson,
  \enquote{Terahertz field–induced ferroelectricity in quantum paraelectric
  {SrTiO3},} {\protect\JournalTitle{Science}} \textbf{364}, 1079--1082 (2019).
  Publisher: American Association for the Advancement of Science.

\bibitem{de_la_torre_colloquium_2021}
A.~de~la Torre, D.~M. Kennes, M.~Claassen, S.~Gerber, J.~W. McIver, and M.~A.
  Sentef, \enquote{Colloquium: {Nonthermal} pathways to ultrafast control in
  quantum materials,} {\protect\JournalTitle{Reviews of Modern Physics}}
  \textbf{93}, 041002 (2021). Publisher: American Physical Society.

\bibitem{sood_electrochemical_2021}
A.~Sood, A.~D. Poletayev, D.~A. Cogswell, P.~M. Csernica, J.~T. Mefford,
  D.~Fraggedakis, M.~F. Toney, A.~M. Lindenberg, M.~Z. Bazant, and W.~C. Chueh,
  \enquote{Electrochemical ion insertion from the atomic to the device scale,}
  {\protect\JournalTitle{Nature Reviews Materials}} \textbf{6}, 847--867
  (2021). Number: 9 Publisher: Nature Publishing Group.

\bibitem{poletayev_persistence_2021}
A.~D. Poletayev, M.~C. Hoffmann, J.~A. Dawson, S.~W. Teitelbaum, M.~Trigo,
  M.~S. Islam, and A.~M. Lindenberg, \enquote{The {Persistence} of {Memory} in
  {Ionic} {Conduction} {Probed} by {Nonlinear} {Optics},}
  {\protect\JournalTitle{arXiv:2110.06522 [cond-mat, physics:physics]}}
  (2021). ArXiv: 2110.06522.

\bibitem{poletayev_defect-driven_2021}
A.~D. Poletayev, J.~A. Dawson, M.~S. Islam, and A.~M. Lindenberg,
  \enquote{Defect-{Driven} {Anomalous} {Transport} in {Fast}-{Ion} {Conducting}
  {Solid} {Electrolytes},} {\protect\JournalTitle{Nature Materials}}  (2022).
  In press, pre-print available at arXiv: 2105.08761.

\bibitem{house_first-cycle_2020}
R.~A. House, G.~J. Rees, M.~A. Pérez-Osorio, J.-J. Marie, E.~Boivin, A.~W.
  Robertson, A.~Nag, M.~Garcia-Fernandez, K.-J. Zhou, and P.~G. Bruce,
  \enquote{First-cycle voltage hysteresis in {Li}-rich 3d cathodes associated
  with molecular {O2} trapped in the bulk,} {\protect\JournalTitle{Nature
  Energy}} \textbf{5}, 777--785 (2020). Number: 10 Publisher: Nature Publishing
  Group.

\bibitem{gent_design_2020}
W.~E. Gent, I.~I. Abate, W.~Yang, L.~F. Nazar, and W.~C. Chueh, \enquote{Design
  {Rules} for {High}-{Valent} {Redox} in {Intercalation} {Electrodes},}
  {\protect\JournalTitle{Joule}} \textbf{4}, 1369--1397 (2020). Publisher:
  Elsevier.

\bibitem{abate_coulombically-stabilized_2021}
I.~I. Abate, C.~D. Pemmaraju, S.~Y. Kim, K.~H. Hsu, S.~Sainio, B.~Moritz,
  J.~Vinson, M.~F. Toney, W.~Yang, W.~E. Gent, T.~P. Devereaux, L.~F. Nazar,
  and W.~C. Chueh, \enquote{Coulombically-stabilized oxygen hole polarons
  enable fully reversible oxygen redox,} {\protect\JournalTitle{Energy \&
  Environmental Science}} \textbf{14}, 4858--4867 (2021). Publisher: The Royal
  Society of Chemistry.

\bibitem{mishra_ultrafast_2015}
P.~K. Mishra, O.~Vendrell, and R.~Santra, \enquote{Ultrafast {Energy}
  {Transfer} from {Solvent} to {Solute} {Induced} by {Subpicosecond} {Highly}
  {Intense} {THz} {Pulses},} {\protect\JournalTitle{The Journal of Physical
  Chemistry B}} \textbf{119}, 8080--8086 (2015). Publisher: American Chemical
  Society.

\bibitem{mishra_prospects_2018}
P.~K. Mishra, V.~Bettaque, O.~Vendrell, R.~Santra, and R.~Welsch,
  \enquote{Prospects of {Using} {High}-{Intensity} {THz} {Pulses} {To} {Induce}
  {Ultrafast} {Temperature}-{Jumps} in {Liquid} {Water},}
  {\protect\JournalTitle{The Journal of Physical Chemistry A}} \textbf{122},
  5211--5222 (2018). Publisher: American Chemical Society.

\bibitem{kas_advanced_2021}
J.~J. Kas, F.~D. Vila, C.~D. Pemmaraju, T.~S. Tan, and J.~J. Rehr,
  \enquote{Advanced calculations of {X}-ray spectroscopies with {FEFF10} and
  {Corvus},} {\protect\JournalTitle{Journal of Synchrotron Radiation}}
  \textbf{28}, 1801--1810 (2021). Number: 6 Publisher: International Union of
  Crystallography.

\bibitem{sorokin_information_2017}
A.~A. Sorokin, S.~V. Makogonov, and S.~P. Korolev, \enquote{The {Information}
  {Infrastructure} for {Collective} {Scientific} {Work} in the {Far} {East} of
  {Russia},} {\protect\JournalTitle{Scientific and Technical Information
  Processing}} \textbf{44}, 302--304 (2017).

\end{thebibliography}


\section{Appendix}
X-ray Absorption Near Edge Spectra (XANES) at the Na K-edge were simulated for selected structural motifs of Na $\beta/\beta^{\prime\prime}$-alumina derived from classical molecular dynamics simulations of the electric field driven ion hopping process. The ab initio real-space multiple scattering method as implemented in the FEFF10\,\cite{kas_advanced_2021} code was employed for this purpose. Accordingly, cluster models of  Na $\beta/\beta^{\prime\prime}$-alumina with ion configurations representative of both the un-perturbed and electric field perturbed systems were considered. In each case the clusters (10×10×10 supercells) consisting of approximately 12000 atoms were constructed centered on the XANES target Na ion of interest. Full multiple scattering (FMS) and self-consistent field (SCF) radii were both set to 9 Å around the reference Na site to ensure convergence in real-space and to satisfy the convergence of spectra less than $10^{-3}$ arb. u. at each point. 

\end{document}